\documentclass[11pt]{article}
\usepackage{amsfonts}
\usepackage{latexsym,amsmath}
\usepackage{amssymb,array}
\usepackage{calc}
\RequirePackage[dvips]{graphicx}
\usepackage{graphics}
\usepackage{tikz}
\usepackage{multicol}
\usepackage{amsmath}
\usepackage{longtable}
\usepackage[numbers]{natbib}
\usepackage{pdflscape}
\usepackage{fdsymbol}
\usepackage{authblk}
\usepackage{float}
\usepackage{verbatim}
\usepackage[colorlinks=true]{hyperref}
\hypersetup{urlcolor=blue, citecolor=red}
\makeatletter \oddsidemargin 0in \evensidemargin 0in \textwidth
16cm\textheight 20cm 
\newtheorem{t1}{Theorem}[section]
\newtheorem{d1}{Definition}[section]

\newtheorem{l1}{Lemma}[section]

\newtheorem{ex1}{Example}[section]

\begin{document}
		\title{Active Redundancy Allocation Strategy at Component and System Level}
	
	\author[1]{Bidhan Modok}
	\author[2]{Shovan Chowdhury\footnote{Corresponding author e-mail:shovanc@iimk.ac.in}}
	\author[1]{Amarjit Kundu}
	
		\affil[1] {Department of Mathematics, Raiganj University, West Bengal, India}
	\affil[2] {Decision Sciences and Operations Management Area, Indian Institute of Management Kozhikode, Kozhikode, India}
	\maketitle
	\begin{abstract}
Researchers and practitioners in the field of reliability engineering and optimization frequently use active redundancy techniques to intensify the performance of systems. In this article, we study allocation strategies of non-matching active redundancies (spares) in coherent systems consisting of possibly dependent and identical components for achieving better system reliability. The dependence of the components is modeled through copulas using the distortion function. Sufficient conditions are derived to establish optimal allocation strategies for two heterogeneous active redundancies at the component or system levels. Moreover, the results are true for the component lifetimes following a general family of parametric distributions. The results guarantee the likelihood ratio (reversed hazard) ordering between the coherent systems at the component level (system level) active redundancies. Some aging properties are also established in this endeavor. Several examples are provided to demonstrate the theoretical results.

\end{abstract}

{\bf Word count}: 8514 words (Approximate).\\
{\bf Keywords and Phrases}: Coherent system, Active redundancy, Stochastic order, Aging class.

	\section{Introduction}
\begin{figure}[ht]
	\centering
	\begin{minipage}[]{0.45\linewidth}
		\includegraphics[height=4.0 cm, width=3.9 cm]{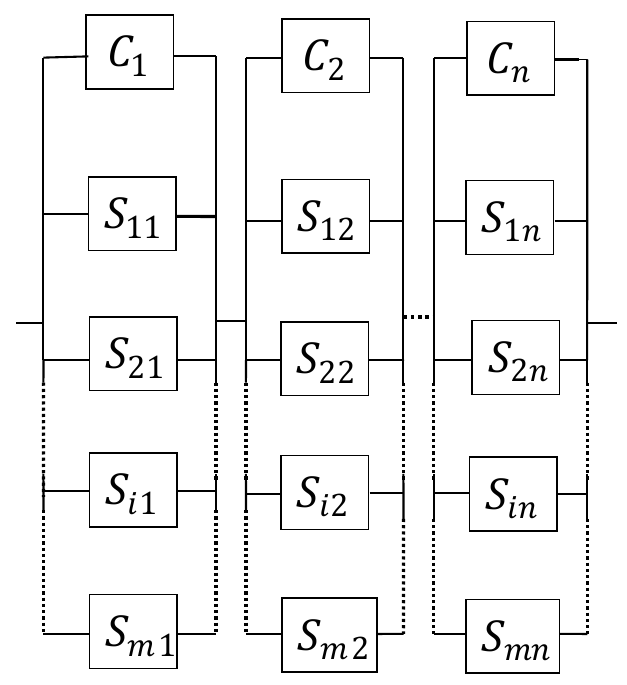}
		\centering{\small{\\$\left(a\right)$ Component level redundant system}}
	\end{minipage}
	\;
	\begin{minipage}[]{0.45\linewidth}
		\includegraphics[height=4.0 cm, width=3.9 cm]{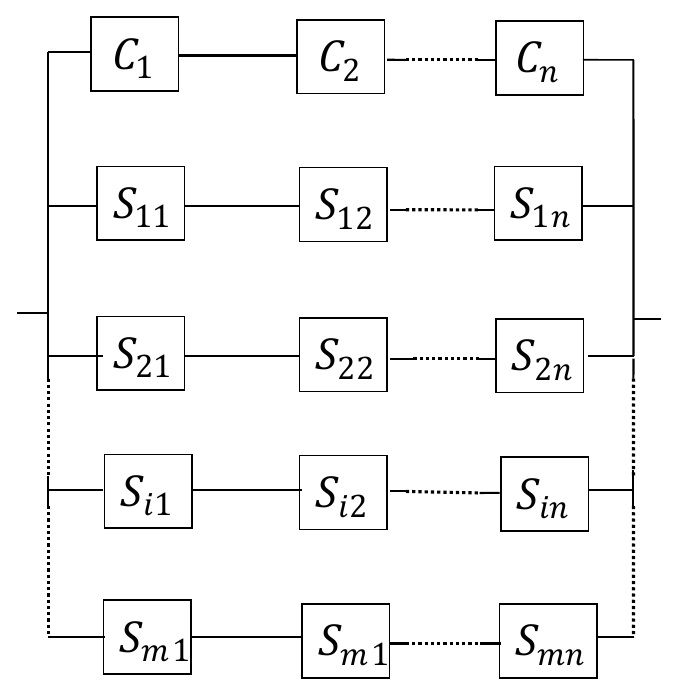}
		\centering{\small{\\$\left(b\right)$ System level redundant system}}
	\end{minipage}\caption{Multiple redundant system}\label{fig0}
\end{figure}

	\setcounter{equation}{0}
\hspace*{0.2in} The use of active redundancy is prevalent in the management of complex systems. One common way of improving the reliability of a coherent system is to allocate active redundancies (or spares) into the system. Recall that a system is said to be coherent if all of its components are relevant and each component’s performance affects the system’s performance (cf.\cite{ba}). The active redundancy technique involves allocating a spare in parallel with an original component so that they work simultaneously when the system is turned on and to facilitate uninterrupted functioning of the system.
In active redundancy, spares may be provided to a coherent system either as component level redundancy or system level redundancy. In component level redundancy, an active spare is provided to each component, while in system level redundancy, the coherent system is duplicated and attached as an active redundant spare to the original system. Let us consider a coherent system consists of $n$ components $C_1,C_2,...C_n$ with $X_1,X_2,...X_n$ being the lifetime of the original components. For active redundancy at component level, suppose $m$ spares $(S_{1j},S_{2j},...S_{mj})$ are assigned to the original component $C_j;~j=1,2,...,n$ in parallel (see Figure \ref{fig0}(a)), to form a system having same structure function as of the original coherent system. Again, for active redundancy at system level, $m$ coherent sub-systems are formed by the components $(S_{i1},S_{i2},...S_{in});~i=1,2,...,m$, and are assigned to the original system $(C_1,C_2,...C_n)$ in parallel (see Figure \ref{fig0}(b)).\\

\hspace*{0.2in} Reliability engineers adopt active redundancy to improve system reliability, particularly in situations where the time between component failure and the recovery time is not permitted. On the other hand, design engineers ensure reliability optimization in terms of adopting optimal allocation policies of a given number of active redundancies to the components of coherent systems. In the past decades, many researchers contributed to the literature on the optimal allocation strategies at component level and system level redundancy. There are two threads of methods used in this context. The first thread revolves around using constrained optimization and algorithm-based methods, while the second thread applies to the stochastic ordering-based method. The current paper aims to use stochastic orders to compare the reliability of two coherent systems at the component or system level redundancies under different allocation strategies and to ascertain the optimum one. In other words, the present work attempts to design an improved reliability system by allocating appropriate redundant components from the available options either at the component or at the system level. \\
\hspace*{0.2in} The landmark paper by \cite{ba} states that the active redundancy at the component level is in general more reliable than the redundancy at the system level in the sense of the usual stochastic ordering, which later became popular as the BP principle and witnessed significant developments over the last four decades. The generalizations of the BP principle can be observed for three distinct types of the component and spare, and is described in the following table.
%\caption{table}{Component and spare types}\label{tab11}
\begin{center}
	{\bf Component and spare types}
\small{\begin{tabular}{|c|c|c} 
\hline
Type & Definition \\
\hline
Complete matching & Components and redundancies are identical \\
Matching  & Coupled redundancies are identical as the original components  \\
Non-matching  & None of the coupled redundancies are identical as the original components  \\
\hline
\end{tabular}}
\end{center}
In the case of complete matching, \cite{bo} and \cite{gu} extended the BP principle to the hazard rate ($hr$) ordering and reversed hazard ($rh$) rate ordering respectively. \cite{si} strengthened the BP principle to the likelihood ratio ($lr$) ordering for the $k$-out-of-$n$ system. The results related to the BP principle under the matching spares can be found in \cite{mi}, \cite{na}, \cite{ha}, \cite{zh}, and the references therein. The case of non-matching spares finds very few papers establishing the BP principle due to the complexity of distribution theory (see \cite{mi}, \cite{ha}). All the research work mentioned so far have assumed component lifetime to be independent of each other. \\  
\hspace*{0.2in} In practical situations, however, there may be a structural dependency among the components of the system. Such dependency may be attributed to a common shock due to various environmental factors such as stress, temperature, voltage, and others, affecting the system’s components. For the dependent and identically distributed ($did$) components of the original system, \cite{gu1} find sufficient conditions for the BP principle to hold, in terms of some stronger stochastic orders for matching and non-matching spares. Later, \cite{da} generalized the BP principle from the way of parallel assembly to that of $k$-out-of-$n$ assembly, which was further generalized for the complete and non-matching spares by \cite{zh1} under multiple redundancies. \cite{ya} investigated conditions for multiple active redundancies at the component versus system level for both matching and non-matching spares. However, the study of dependent and non-identically (heterogeneous) distributed ($dnid$) components is rarely found in the literature due to the complexity of the distribution theory.\\
\hspace*{0.2in}  Interestingly, comparisons of two coherent systems at the component level or the system level active redundancy have been rarely studied. \cite{ke} first compared two coherent systems of $did$ components with non-matching active redundancies with lifetimes of both components and spares following the proportional hazards or proportional reversed hazard model. Recently, \cite{pa, pa2} investigated similar systems with scale and proportional odds family of distributions. These studies are extremely important from the lens of optimum selection of non-matching redundant components which can result in an improved system reliability.\\
\hspace*{0.2in} The present work is motivated by the following example. It is well known that redundant cooling system plays a significant role to protect the primary cooling system and ensure uninterrupted operations of electronic devices, data centers, telecommunications network etc. It enables continuous cooling even if one cooling component experiences a failure resulting in service excellence. Suppose we have a cooling system consists of four cooling components which are identical in nature, following any parametric lifetime distribution, and also dependent on each other through copula. Assume that the system follows $X_{2:4}$ (2-out-of-4) coherent structure. It is practical to assume the spares to be non-identical (non-matching) as the cooling components may be manufactured by different firms. Suppose five different brands $\left(S_1,S_2\ldots, S_5\right)$ of cooling components are available in the market with different failure rates. In such situation, an interesting problem is to select redundant components optimally. Assume that there are two 2-out-of-4 identical coherent systems, and, two sets of active redundant components for the two systems are $(S_1, S_3)$, and $(S_2,S_5)$ respectively. The natural question is which system of active redundancy at the component level is more reliable?\\     
\hspace*{0.2in} The contribution of the paper is multi-fold. First, we compare reliability of two coherent systems with $did$ components and active multiple non-matching spares either at component level or at system level assuming a general class of lifetime distribution functions. In this sense, the results are more general than the few existing ones. Second, sufficient conditions are provided to guarantee the $lr$ ordering between the coherent systems at component level active redundancy. Third, sufficient conditions are provided to guarantee the $rh$ ordering between the coherent systems at system level active redundancy. Fourth, some aging class properties are guaranteed for both component and system level active redundancies. No existing work in this domain has investigated the aging properties to the best of our knowledge.\\
\hspace*{0.3in} The paper is organized as follows. Section 2 discusses some preliminary concepts related to the present work. Section 3 and Section 4 derive results for coherent systems with active redundancy at the component level and system level respectively. The notations $g'(u)$ and $g''(u)$ denote, respectively, the first and second order derivative of the function $g(u)$ with respect to $u$ throughout the paper.
	
\section {Preliminaries}
	\setcounter{equation}{0}
\hspace*{0.3in}
In this section, we furnish some important concepts through their definitions which will be used in the next sections. Throughout the paper, the word increasing (resp. decreasing) and nondecreasing (resp. nonincreasing) are used interchangeably, and $\mathbb{R}$ denotes the set of real numbers $\{x:-\infty<x<\infty\}$. Consider an absolutely continuous random variable $X$, where $f_X(\cdot)$ denotes its probability density function (pdf), $F_X(\cdot)$ its cumulative distribution function (cdf), $h_X(\cdot)$ its hazard rate function, and $\tilde{h}_X(\cdot)$ its reversed hazard rate function. The survival or reliability function (sf) of $X$ can be represented as $\overline{F}_X(\cdot) = 1 - F_X(\cdot)$.

The following definitions on majorization based stochastic inequalities can be obtained from  \cite{ma}.

	\begin{d1} Let ${\bf x} = (x_1, x_2, \dots, x_n) \in \mathbb{R}^n$ and ${\bf y} = (y_1, y_2, \dots, y_n) \in \mathbb{R}^n$ represent two arbitrary vectors. Let $x_{(1)}\leq x_{(2)}\leq\ldots \leq x_{(n)}$ be an ascending arrangement of $\bf{x}\in\mathbb{R}^n. $
	\begin{enumerate}
		\item The vector $\bf{x}$ is considered to majorize the vector $\bf{y}$ (noted as $\bf{x} \succeq_m \bf{y}$) if
		\[
		\sum_{i=1}^{j} x_{(i)} \leq \sum_{i=1}^{j} y_{(i)}, \quad \text{for } j = 1, 2, \dots, n - 1,
		\]
		along with
		\[
		\sum_{i=1}^{n} x_{(i)} = \sum_{i=1}^{n} y_{(i)}.
		\]
		\item The vector $\bf{x}$ is said to weakly supermajorize the vector $\bf{y}$ (denoted as $\bf{x} \succeq^w \bf{y}$) if
		\[
		\sum_{i=1}^{j} x_{(i)} \leq \sum_{i=1}^{j} y_{(i)}, \quad \text{for } j = 1, 2, \dots, n.
		\]
	%	\item Similarly, $\bf{x}$ is said to weakly submajorize $\bf{y}$ (noted as $\bf{x} \succeq_w \bf{y}$) if
	%	\[
	%	\sum_{i=j}^{n} x_{(i)} \geq \sum_{i=j}^{n} y_{(i)}, \quad \text{for } j = 1, 2, \dots, n.
	%	\]
	\end{enumerate}
	\end{d1}
%It is easy to show that $\mathbf{x}\stackrel{\rm m}{\succeq}\mathbf{y}\Rightarrow\mathbf{x}\stackrel{\rm w}{\succeq} \mathbf{y}.$\\
In order to make comparisons among various order statistics, researchers use stochastic order principles. Over time, different type of stochastic orders have been proposed and analyzed in the literature. While detailed explanations, motivations, and applications of these orders can be found in \cite{shak1}, we present some definitions below for the sake of completeness and reader accessibility.

\begin{d1} Let $X$ and $Y$ be two absolutely continuous random variables with respective supports $(l_X, u_X)$ and $(l_Y, u_Y)$, where $u_X$ and $u_Y$ may be positive infinity, and $l_X$ and $l_Y$ may be negative infinity. Then, $X$ is said to be smaller than $Y$ in $i)$ likelihood ratio ($lr$) order, denoted as $X \leq_{lr} Y$, if $\frac{f_Y(t)}{f_X(t)}$ is increasing in $t \in (l_X, u_X) \cup (l_Y, u_Y);$  $ii)$ hazard rate ($hr$) order, denoted as $X \leq_{hr} Y$, if $\bar{F}_Y(t) / \bar{F}_X(t)$ is increasing in $t \in (-\infty, \max(u_X, u_Y));$  $iii)$ reversed hazard rate (rh) order, denoted as $X \leq_{rh} Y$, if $\frac{F_Y(t)}{F_X(t)}$ is increasing in $t \in (\min(l_X, l_Y), \infty);$  $iv)$ usual stochastic ($st$) order, denoted as $X \leq_{st} Y$, if $\bar{F}_X(t) \leq \bar{F}_Y(t)$ for all $t \in \mathbb{R}$ .
\end{d1}
%In the following diagram, we present a chain of implications of the stochastic orders.
%\vspace{0.17 in}
%\\\hspace*{1.7 in}$~~~~~~X\leq_{hr}Y$
%\\\hspace*{1.7 in}$~~~~~~~~~~~\uparrow ~~~~~~~\searrow$
%\\\hspace{6 in} $~~~~~~~~~~~~~~~~~~~~~~~~~~~~~~~~~~~~~~~~~X\leq_{lr}Y~~\rightarrow~~X\leq_{st}Y.$

%\hspace{2.51 cm}$~~~~~~~~~~~~~~~~~~~~~~~~~\downarrow~~~~~~~~~\nearrow$

%\hspace{2 cm}$~~~~~~~~~~~~~~~~~~~~~~~~X\leq_{rh}Y$
 	 
\hspace*{0.3 in} Now, let us recall that a copula associated with a multivariate distribution function $F$ is a function $C:\left[0,1\right]^n\longmapsto\left[0,1\right]$ satisfying: $F(\mathbf{x})=C\left(F_{1}(x_1),..., F_{n}(x_n)\right),$ where the $F_i$'s, $1\leq i\leq n$ are the univariate marginal distribution functions of $X_i$'s. Similarly, a survival copula associated with a multivariate survival function $\overline{F}$ is a function $\overline{C}:\left[0,1\right]^n\longmapsto\left[0,1\right]$ satisfying:
$$\overline{F}(\mathbf{x})=P\left(X_1>x_1,...,X_n>x_n\right)=\overline{C}\left(\overline{F}_1(x_1),...,\overline{F}_n(x_n)\right),$$ 
where, for $1\leq i\leq n$, $\overline{F}_i(\cdot)=1-F_i(\cdot)$  are the univariate survival functions. %In particular, a copula $C$ is Archimedean if there exists a generator $\psi:\left[0,\infty\right]\longmapsto\left[0,1\right]$ such that
%$$C\left(\mathbf{u}\right)=\psi\left(\psi^{-1}(u_1)+...+\psi^{-1}(u_d)\right).$$
For more detail on Archimedean copula, see \cite{ne} and \cite{mc}. In this paper, we have used Clayton and Gumbel families of Archimedean copulas and FGM copula as well, as given by 
\begin{itemize}
	\item [(i)] Clayton: $C(u_1,u_2,\cdots, u_n:\theta)=\left(\sum_{i=1}^n  u_i^{-\theta}-(n-1)\right)^{-\frac{1}{\theta}}, \; \theta \in [-1,\infty)\setminus{\{0\}}.$
	\item[(ii)] Gumbel: $C(u_1,u_2,\cdots, u_n:\theta)= exp\left({-\left[\sum_{i=1}^n(-\ln u_i)^{\theta}\right]^{\frac{1}{\theta}}}\right), \; \theta \in [1,\infty).$
	\item[(iii)] FGM: $C(u_1,u_2,\cdots, u_n:\theta)=\prod_{i=1}^nu_i\left(1+\theta\prod_{i=1}^n(1-u_i)\right),  \; \theta \in [-1,1].$
\end{itemize}
	\section {Allocation of Active Redundancy at Component Level}\label{sec3}
	\setcounter{equation}{0}
	\hspace*{0.3 in} In this section two systems, having component level redundancies, as described in the introduction section (see Figure~\ref{fig0} (a)), are compared with respect to different stochastic orderings with examples. In this context, some results are derived which can help design more reliable systems by allocating appropriate redundant components from the available spares.\\
\cite{na1} obtained a representation of the system reliability ($\bar{F}_{T}$) for $did$ components as a distorted function of the component reliability $(\bar{F})$ given by $\bar{F}_{T}\equiv q_\theta(\bar{F})$ where $q_\theta:[0,1]\rightarrow [0,1]$ is a continuous increasing function such that $q_\theta(0)=0$ and $q_\theta(1)=1$ and depends on the system structure and the survival copula of the joint distribution of the component lifetimes. The parameter $\theta$ assumes the role of the dependence parameter of the copula function associated with the system. For more details on the distorted functions, their developments and applications, readers are referred to \cite{lin}, \cite{nav}, and the references therein. Thus, the reliability function of a coherent system at component level redundancy is given as 
	\begin{equation}\label{eqs0}
		\bar{F}_{c}(t)=q_\theta\left(1-\prod_{i=0}^mF_i(t)\right),
	\end{equation}
where $F_i(t)$ is the distribution function of the $i^{th}$ component.\\
	Now, suppose there are two coherent systems, both of which are composed of $n$ number of $did$ original components $C_i$ and $C_i^*$, $i=1,2,\ldots, n$, having cdfs and distortion functions ($F\left(t,b_0\right),$ $q_{\theta_1}(\cdot)$) and ($F\left(t,b^*_0\right),$ $q_{\theta_2}(\cdot)$) respectively. Also let both the systems have active redundancy at component level with $m$ number of non-matching redundant components. For $i=1,2,\ldots, n$, let $j^{th}$ redundant component allocated to each of $C_i$ ($C_i^*$) has the cdf $F(t,b_j)$ ($F(t,b_j^*)$), $j=1,2,\ldots,m$ (see Figure~\ref{fig0} (a)). It is to be mentioned here that $b_j$, $b_j^* \in \mathbb{R},$~$j=0,1,2\ldots, m$ are the parameters of the distribution function $F$. Now, let the random variables (rvs) $X_c$ and $X_c^*$ represent lifetimes of these two systems. Thus, following ($\ref{eqs0})$, the reliability functions of these two systems are given respectively by 
	\begin{equation}\label{eqs1}
	\bar{F}_c(t)=q_{\theta_1}\left(1-\prod_{j=0}^mF(t;b_j)\right)\ {\text and }\ \bar{F}_c^*(t)=q_{\theta_2}\left(1-\prod_{j=0}^m F(t;b^*_j)\right),
	\end{equation}
	where $F(\cdot)$ is the baseline distribution function. In this section, depending upon various conditions, the lifetimes $X_c$ and $X_c^*$ of the two systems are compared with respect to various stochastic orderings. Conceptually, the sufficient conditions presented in the following results capture intuitive physical behaviors of redundant components and dependence structures of the original components, thereby guiding design engineers in identifying the optimal allocation strategy for available active redundancies at the component level. For instance, the monotonicity of the distortion function $q_{\theta}(u)$ ensures that the dependence among components constituting the system, as captured by copulas, varies consistently with the dependence parameter $\theta$. In practice, such a condition is mild and is readily satisfied by widely used copula families such as Clayton, Gumbel, and FGM. The condition of log-concavity or log-convexity of $F(t,b)$ is a characteristic feature of some lifetime distributions used in reliability modeling. For notational convenience, the systems are represented by $\left(F, \boldsymbol{b}, q_{\theta_1}\right)$ and $\left(F, {\boldsymbol b}^*, q_{\theta_2}\right)$ respectively with ${\boldsymbol b}=\left(b_0,b_1,\ldots, b_m\right)$ and ${\boldsymbol b}^*=\left(b_0^*,b_1^*,\ldots, b_m^*\right)$. Let the first and second order derivative of $F$ with respect to each of $b_i$ and $b_i^*$ exists. The definitions of Schur-convex and Schur-concave functions are given below which will be used to prove the upcoming theorems.\\
		\begin{d1} Let $I \subseteq \mathbb{R}$. A function $\psi : I^n \rightarrow \mathbb{R}$ is said to be Schur-convex (resp. Schur-concave) on $I^n$ if $x \preceq_m y$ implies $\psi(x) \leq$ (resp. $\geq$) $\psi(y)$ for all $x, y \in I^n$.
		\end{d1}
		The following lemma is used to prove some of the upcoming theorems, the proof of which  is very straight forward and hence omitted. The proofs of the theorems in Section 3 and Section 4 can be found in the APPENDIX.
		\begin{l1}\label{l1}
			
			Let $I \subset \mathbb{R}$ be an interval, and let $g : I \to \mathbb{R}$ be a concave function. Define  
			\[
			\phi(\boldsymbol{x}) = \sum_{i=1}^n g(x_i),
			\]
			where $\boldsymbol{x} = (x_1, x_2, \ldots, x_n) \in I^n$. Then, $\phi(\boldsymbol{x})$ is Schur-concave on $I^n$, and consequently if $\boldsymbol{x} \preceq_m \boldsymbol{y}$ on $I^n$  then 
				$
				\phi(\boldsymbol{x}) \geq \phi(\boldsymbol{y}).
				$
			
		\end{l1}
		
\begin{figure}[H]
	%\centering
	\begin{minipage}[]{0.48\linewidth}
		\includegraphics[height=4.0 cm, width=7.9 cm]{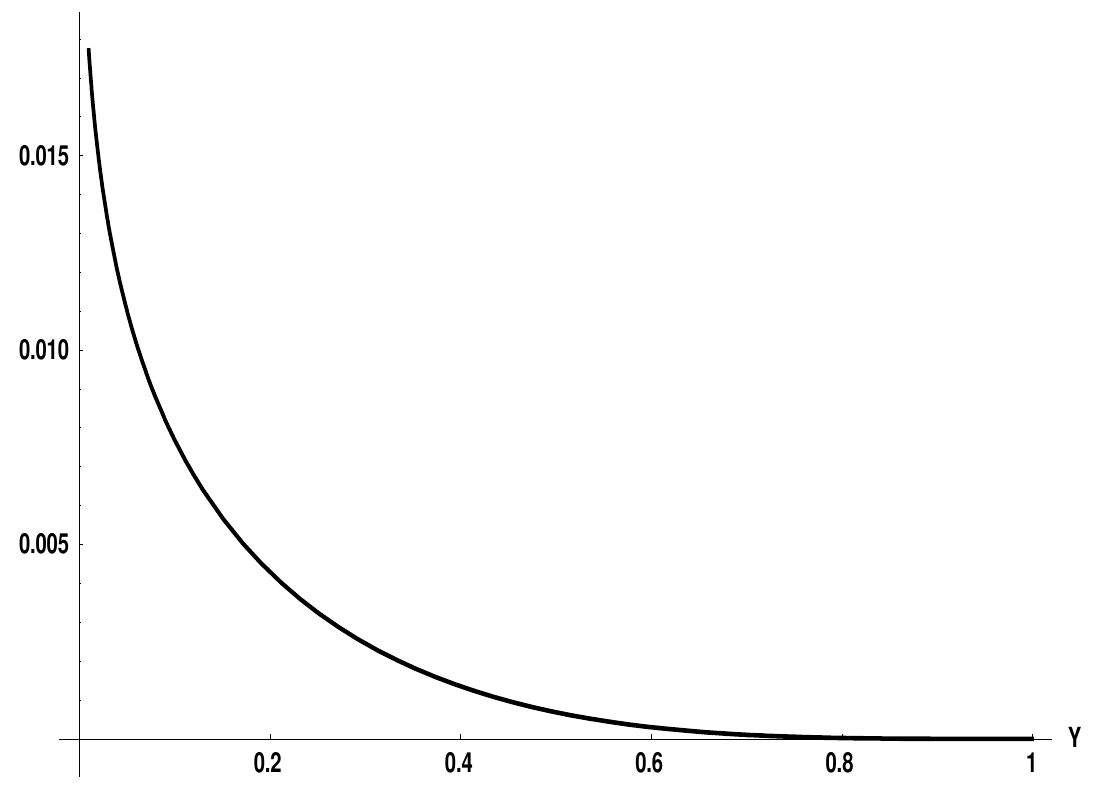}\\
		\centering{\small{$\left(a\right)$ Curve of $\bar{F}_c^*(y)-\bar{F}_c(y)$}}
	\end{minipage}
	\quad
	\begin{minipage}[]{0.48\linewidth}
		\includegraphics[height=4.0 cm, width=7.9 cm]{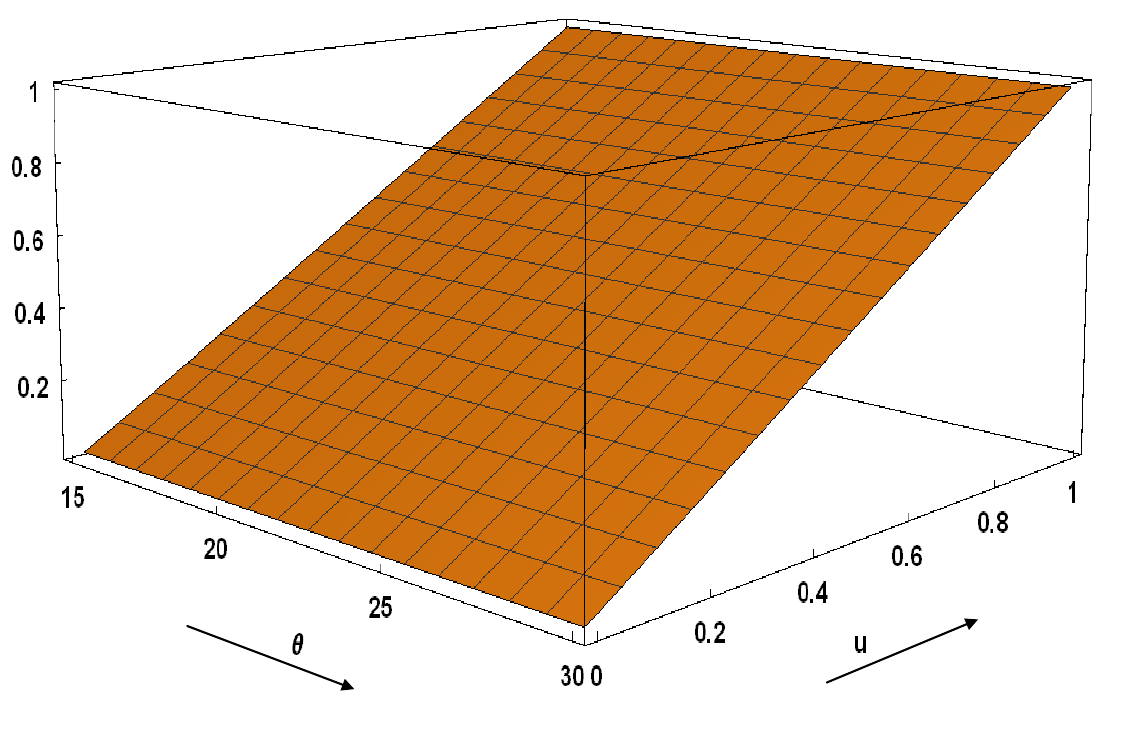}\\
		\centering{\small{ $\left(b\right)$ Curve of $q_{\theta}(u)$}}
	\end{minipage}\caption{Curves for Example \ref{ex3.1}}\label{fig1}
\end{figure}

%In Theorem \ref{th1}, given below, some sufficient conditions are derived for which the survival function of one coherent system for a specific choice of component level redundancy (CLR) is larger than the same system with another choice of CLR. The theorem guarantees that the majorized parameter vector of one set of CLR leads to larger systems lifetime. Hence, Theorem \ref{th1} can help design engineer to allocate an optimal set of redundant components from available options which can improve system reliability. Theorem \ref{th2} can be interpreted in a similar way.
	\begin{t1}\label{th1}
		Let us consider two coherent systems with redundancy at component level having model $\left(F, \boldsymbol{b},q_{\theta_1}\right)$ with lifetime $X_c$ and model $\left(F,\boldsymbol{b^*},q_{\theta_2}\right)$ with lifetime $X_c^*$. For $\theta_1\leq\theta_2$ and for all $u\geq 0$ if
		\begin{itemize}
			\item [i)] $q_{\theta}(u)$ is increasing in $\theta$ and
			\item [ii)] $F(t;b)$ is increasing and log-concave in $b$,
		\end{itemize}
		then $\boldsymbol{b}\preceq^w \boldsymbol{b^*}$ implies $X_c\leq_{st} X_c^*$.
	\end{t1}
Conceptually, Theorem \ref{th1} implies that when two systems share the same structural and distributional form, but differ in the strength of dependence among components (through $\theta$) and the quality of redundant components (through $b_i$, $i=1,2,\cdots,m$), the system with stronger dependence (larger $\theta$) and a majorizing parameter vector (less dispersed allocation of component strengths) achieves larger system lifetime. The theorem holds for redundant components following any distribution function satisfying condition $ii).$ 
	The following example illustrates Theorem $\ref{th1}$ and highlights the practical implications of its sufficient conditions.
	\begin{ex1}\label{ex3.1}
		Let us consider a $3$-out-of-$4$ coherent system having components depending on each other by Gumbel copula. Then form Figure~$\ref{fig1}(b)$ it can be said that, for all $0\leq u\leq 1$, $q_{\theta}(u)=4e^{-(3(-\ln u)^{\theta})^{\frac{1}{\theta}}}-3e^{-(4(-\ln u)^{\theta})^{\frac{1}{\theta}}}$  where $\theta \in [15, \infty)$, is increasing in $\theta.$ Consider the Pareto Type II (Lomax) distribution with cdf $F(x;b)=1-(1+x)^{-b}, \; x\geq0,\; b>0.$ It can be verified that $F(x;b)$ satisfies condition $(ii)$ of Theorem~\ref{th1}. So considering $\boldsymbol{b}=(b_0,b_1,b_2,b_3)=(1.2,0.5,0.4,0.2)$ and $\boldsymbol{b^*}=(b_0^*,b_1^*,b_2^*,b_3^*)=(1,0.5,0.3,0.2)$ giving  $\boldsymbol{b}\preceq^w \boldsymbol{b^*}$ and taking $\theta_1=20$ and $\theta_2=25$  from Figure $\ref{fig1}(a)$, it can be said that for all $x\geq 0,$ $\bar{F}_c(x)\leq \bar{F}_c^*(x),$ showing the result. The substitution $x=-\ln y$, $0<y<1$, is used here to plot the curve.
					\end{ex1}
					
\begin{figure}[H]
		%\centering
		\begin{minipage}[]{0.48\linewidth}
			\includegraphics[height=4.0 cm, width=7.9 cm]{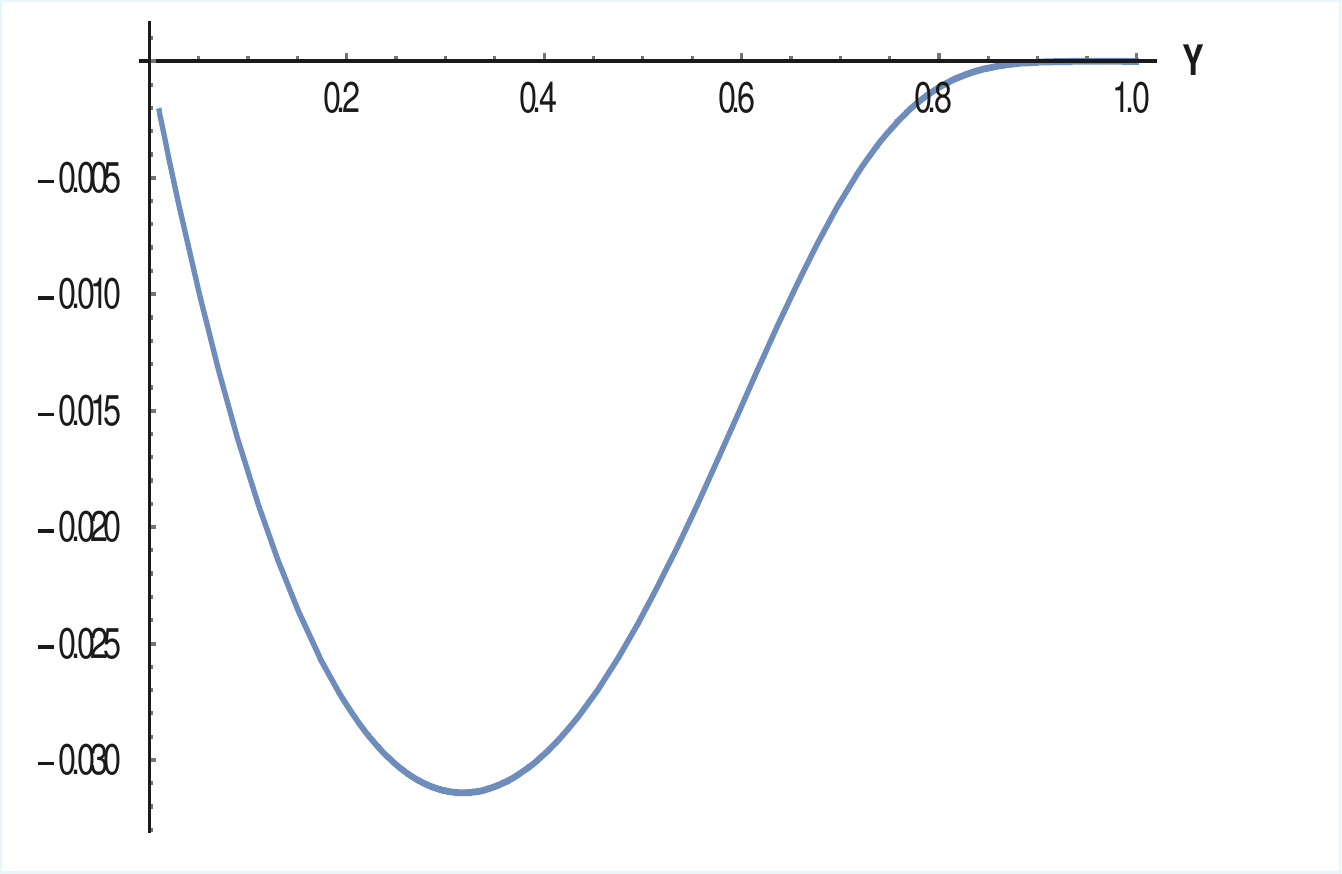}\\
			\centering{\small{$\left(a\right)$ Curve of $\bar{F}_c^*(y)-\bar{F}_c(y)$}}
		\end{minipage}
		\quad
		\begin{minipage}[]{0.48\linewidth}
			\includegraphics[height=4.0 cm, width=7.9 cm]{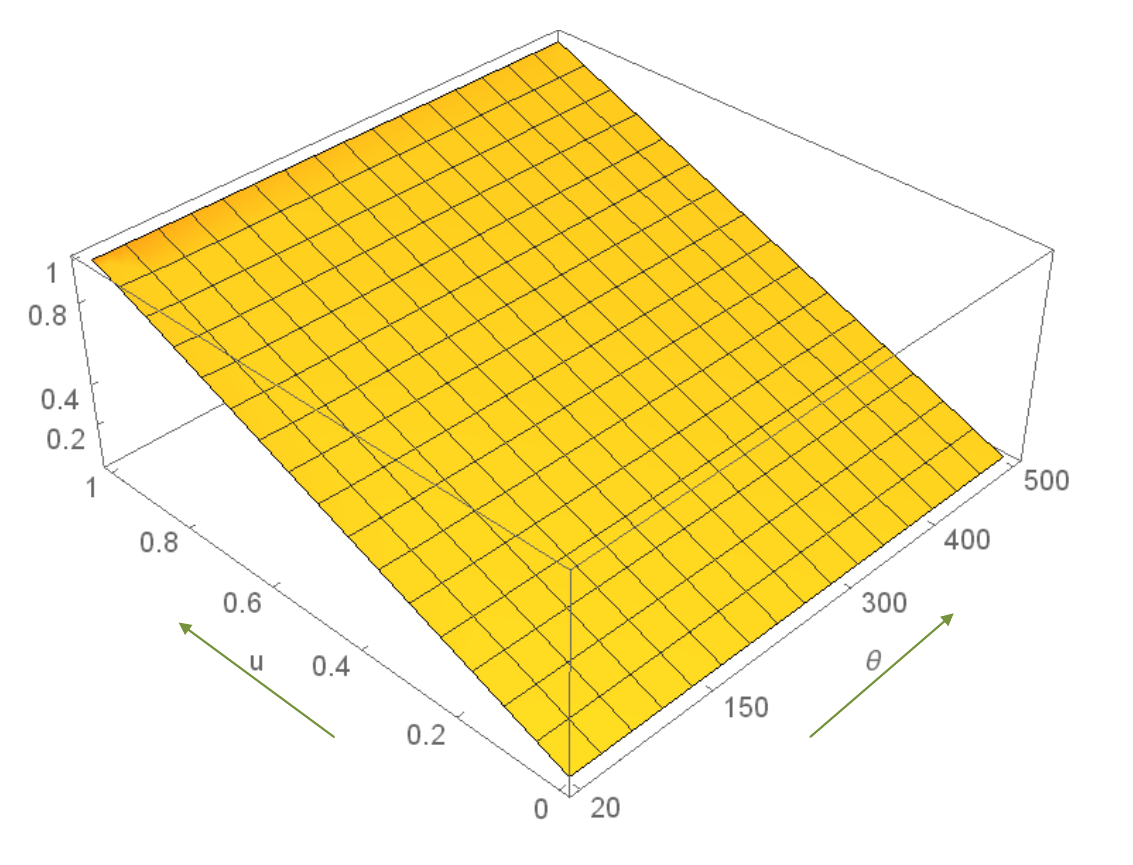}\\
			\centering{\small{ $\left(b\right)$ Curve of $q_{\theta}(u)$}}
		\end{minipage}\caption{Curves for Example \ref{ex3.2}}\label{fig3.2}
	\end{figure}

	The next theorem shows that the same ordering holds between $X_c$ and $X_c^*$, but in reverse order, under different conditions of $q_{\theta}(u)$ and $F(t,b)$. The conceptual interpretation of these sufficient conditions, as well as the structure of the proof, is analogous to that of Theorem $\ref{th1}$ and is therefore omitted here.
	\begin{t1}\label{th2}
		Let us consider two coherent systems with redundancy at component level having model $\left(F, \boldsymbol{b},q_{\theta_1}\right)$ with lifetime $X_c$ and model $\left(F,\boldsymbol{b^*},q_{\theta_2}\right)$ with lifetime $X_c^*$. For $\theta_1\leq\theta_2$ and for all $u\geq 0$ if
		\begin{itemize}
			\item [i)] $q_{\theta}(u)$ is decreasing in $\theta$ and
			\item [ii)] $F(t;b)$ is decreasing and log-convex in $b$,
		\end{itemize}
		then $\boldsymbol{b}\preceq^w \boldsymbol{b^*}$ implies $X_c\geq_{st} X_c^*$.
	\end{t1}
		\begin{ex1}\label{ex3.2}
		To justify Theorem~\ref{th2}, consider the generalized exponential distribution defined by $F(x; b) = \left( 1-e^{-x}  \right)^{\sqrt{b}}$ for $x \geq 0 ,\; b \geq 0$, and consider a coherent system defined as $max(X_1,\min(X_2,X_3,X_4))$, where the components are dependent on each other through the Clayton copula, then  from Figure~$\ref{fig3.2}(b)$ it is observed that the distortion function $q_{\theta}(u) = u + (3u^{-\theta} - 2)^{-1/\theta} - (4u^{-\theta} - 3)^{-1/\theta}$ with $\theta \in [2, \infty) $, is decreasing in $\theta$. It can be verified that $F(x;b)$ satisfies condition $(ii)$ of Theorem~\ref{th2}.
		So considering $\boldsymbol{b}=(b_0,b_1,b_2,b_3)=(1.2,0.5,0.4,0.2)$ and $\boldsymbol{b^*}=(b_0^*,b_1^*,b_2^*,b_3^*)=(1,0.5,0.3,0.2)$ giving  $\boldsymbol{b}\preceq^w \boldsymbol{b^*}$ and taking $\theta_1=3$ and $\theta_2=4$  from Figure $\ref{fig3.2}(a)$, it can be said that for all $x\geq 0,$    $\bar{F}_c(x)\geq \bar{F}_c^*(x),$ showing the result. Here substitution $x=-\ln y$, $0<y<1$, is used, while plotting the curve.
	\end{ex1}

		%\begin{r1}
		%\textcolor{blue}{It is to be mentioned here that the condition that $q_\theta (u)$ is increasing in $\theta$ ensures stronger positive dependence between the components constituting that system. Other than  $q_\theta (u)$ mentioned in Example \ref{ex3.1}, the same function of a $2$-out-of-$3$ coherent system having components depending on each other by Gumbel copula or of a $3$-out-of-$4$ coherent system with Gumbel-Barnett copula possess the same property.\\
	%	On the other hand, for Theorem \ref{th2}, in excess of  Example \ref{ex3.2}, the $q_\theta (u)$ of a $2$-out-of-$3$ coherent system having components depending on each other by FGM copula is also decreasing in $\theta$.}
	%\end{r1}	
	%\begin{comment}
	%\begin{ex1}
	%\textcolor{red}{To justify Theorem~\ref{th2}, consider the \textcolor{blue}{generalized exponential distribution defined by} $F(x; b) = \left( 1-e^{-x}  \right)^{\sqrt{b}}$ for $x \geq 0 ,\; b \geq 0$, and $q_{\theta}(u) = u + (3u^{-\theta} - 2) - (4u^{-\theta} - 3)^{-1/\theta}$ with $\theta \in [-1, \infty) \setminus \{0\}$, which satisfy all the conditions of the theorem.}
		
	%\end{ex1}
	%\end{comment}
	Theorems $\ref{th3}$ and $\ref{th4}$ demonstrate sufficient conditions for $hr$ ordering between $X_c$ and $X_c^*$ and determine the best allocation choice of redundancies. The $hr$ ordering helps compare two systems conditioned on the event that the systems have survived upon a specified time point; in other words, the failure (hazard) rates of two used systems can be compared using $hr$ ordering. In this sense, Theorems \ref{th3} and \ref{th4} indicate that when two systems share the same structural, and distributional form with specific characteristic of reversed hazard rate function, but differ in the strength of dependence among components (through $\theta$) and the quality of redundant components (through $b_i$, $i=1,2,\cdots,m$), one system performs better than the other. Condition $iii)$ confirms that $q_{\theta}(u)$ is log-concave (log-convex) in $\theta$ reflecting consistent dependence sensitivity. Examples \ref{ex3.3} and \ref{ex3.4} demonstrate the sufficient conditions and establish the allocation strategy.

	\begin{t1}\label{th3}
		Let us consider two coherent systems with redundancy at component level having model $\left(F, \boldsymbol{b},q_{\theta_1}\right)$ with lifetime $X_c$ and model $\left(F,\boldsymbol{b^*},q_{\theta_2}\right)$ with lifetime $X_c^*$. For $\theta_1\leq\theta_2$ and for all $u\geq 0$ if 
		\begin{itemize}
			\item [i)] $F(t;b)$ is log-concave in $b$,
			\item [ii)] $\tilde{h}(t;b)$ is  concave (convex) in $b$ and
			\item [iii)] $\frac{(1-u)q_{\theta}'(u)}{q_{\theta}(u)}$ is decreasing (increasing) in $u$ and decreasing (increasing) in $\theta$
		\end{itemize}
		then $\boldsymbol{b}\preceq_m \boldsymbol{b^*}$ implies $X_c\leq_{hr}(\geq_{hr})X_c^*.$ 
	\end{t1}
	
\begin{figure}[H]
			%\centering
			\begin{minipage}[]{0.48\linewidth}
				\includegraphics[height=4.0 cm, width=7.9 cm]{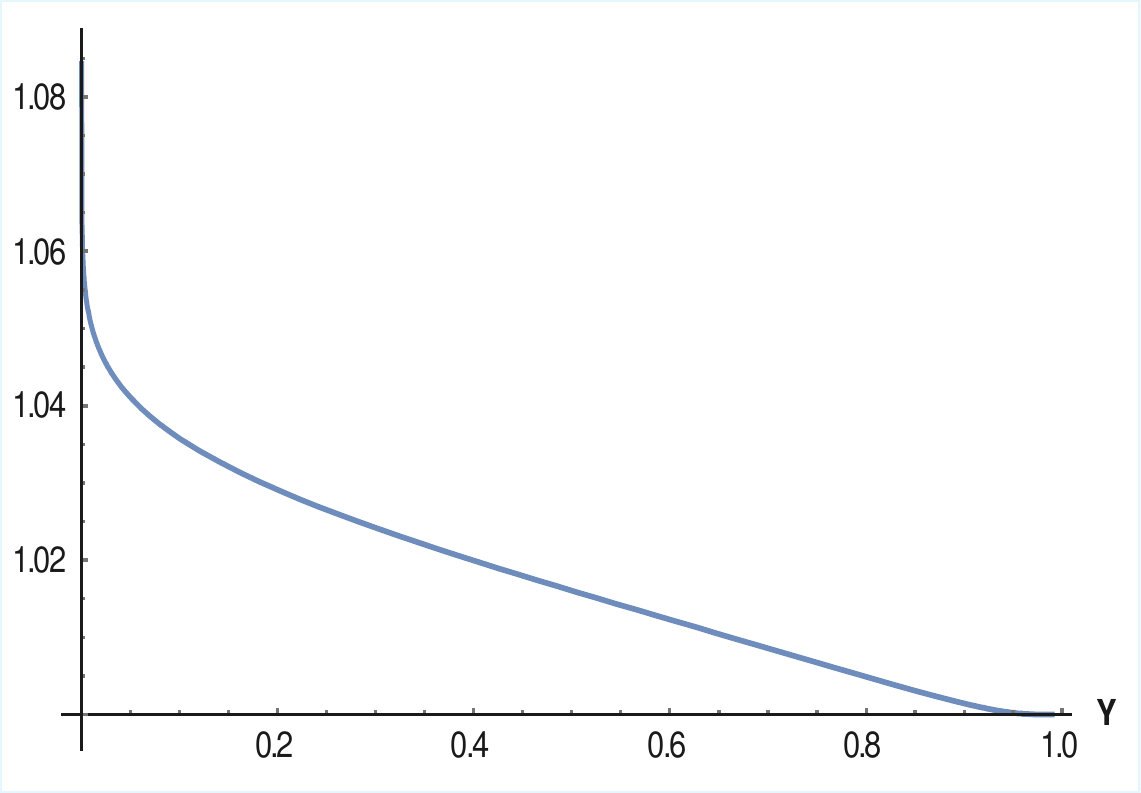}\\
				\centering{\small{$\left(a\right)$ Curve of$\frac{\bar{F}_c^*(y)}{\bar{F}_c(y)}$}}
			\end{minipage}
			\quad
			\begin{minipage}[]{0.48\linewidth}
				\includegraphics[height=4.0 cm, width=7.9 cm]{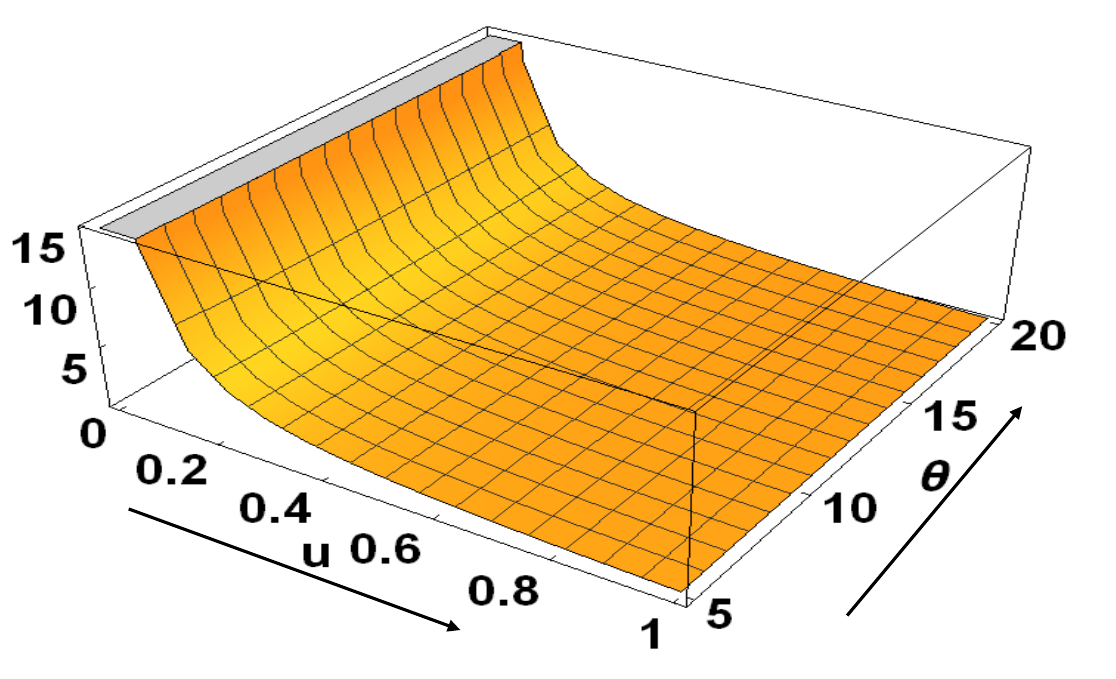}\\
				\centering{\small{ $\left(b\right)$ Curve of $\frac{(1-u)q'_{\theta}(u)}{q_{\theta}(u)}$}}
			\end{minipage}\caption{Curves for Example \ref{ex3.3}}\label{fig2}
		\end{figure}

	%Hence $\xi(t)$ is increasing in $t$, Therefore $$X_c\leq_{hr}X_c^*$$
		\begin{ex1}\label{ex3.3}
		Let us consider a $3$-out-of-$4$ coherent system with component level redundancy dependents on each by Gumbel copula.  Thus the distortion function of the system is given by $q_{\theta}(u)=4e^{-(3(-\ln u)^{\theta})^{\frac{1}{\theta}}}-3e^{-(4(-\ln u)^{\theta})^{\frac{1}{\theta}}}, \theta \in [1,\infty).$ From Figure~$\ref{fig2}(b)$, it can be said that  $\frac{(1-u)q'_{\theta}(u)}{q_{\theta}(u)}$ is decreasing in $u$ and $\theta \in [5,20].$ Now let us consider the inverted exponential distribution defined by $F(x;b)=e^{-\frac{b}{x}},\; b>0, \; x>0$, which satisfy conditions $(i)$ and $(ii)$ of Theorem $\ref{th3}.$
		Now consider $\boldsymbol{b}=(b_0,b_1,b_2,b_3)=(0.05,0.05,0.04,0.02)$ and $\boldsymbol{b^*}=(b_0^*,b_1^*,b_2^*,b_3^*)=(0.06,0.05,0.03,0.02)$ showing that $\boldsymbol{b}\preceq_m \boldsymbol{b^*}.$ Considering $\theta_1=6$ and $\theta_2=7$ Thus Figure $\ref{fig2}(a)$  shows that $X_c$ is smaller than $X_c^*$ in $hr$ ordering. Here substitution $x=-\ln y$, $0<y<1$, is used to capture whole real line.
			\end{ex1}
		
\begin{figure}[H]
	\centering
	\includegraphics[height=4.0 cm, width=7.9 cm]{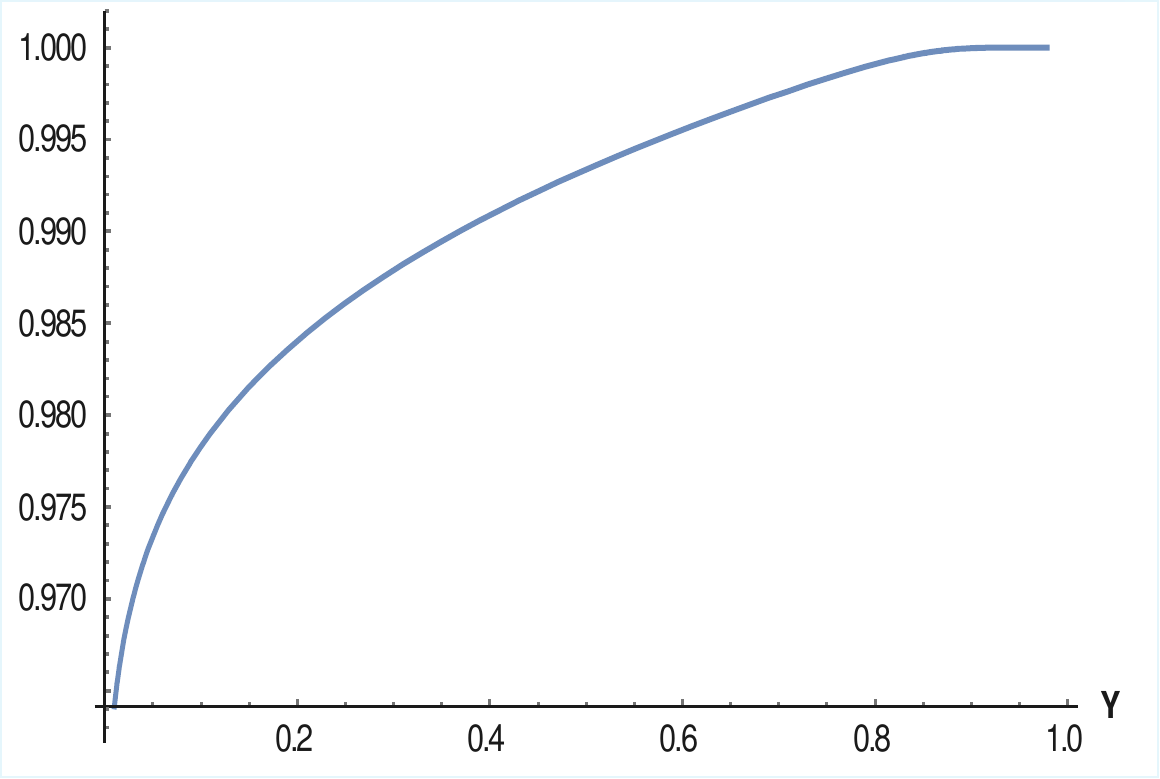}
	\caption{Curve of $\frac{\bar{F}_c^*(y)}{\bar{F}_c(y)}$}\label{Fig3.4}
\end{figure}

	 The proof of Theorem $\ref{th4}$ is similar to Theorem $\ref{th4},$ and hence is omitted. 
	\begin{t1}\label{th4}
		Let $\left(F, \boldsymbol{b},q_{\theta_1}\right)$ and $\left(F,\boldsymbol{b^*},q_{\theta_2}\right),$ described above, be two coherent systems  with lifetimes $X_c$ and $X_c^*$  respectively. Now if $\theta_1\geq\theta_2$ and for all $u\geq 0$,  
		\begin{itemize}
		\item [i)] $F(t;b)$ is log-convex in $b$,
		\item [ii)] $\tilde{h}(t;b)$ is  convex (concave) in $b$ and 
		\item [iii)] $\frac{(1-u)q_{\theta}'(u)}{q_{\theta}(u)}$ is decreasing (increasing) in $u$ and decreasing (increasing) in $\theta$
		\end{itemize}
		then $\boldsymbol{b}\preceq_m \boldsymbol{b^*}$ implies $X_c\geq_{hr} (\leq_{hr})X_c^*.$
	\end{t1}
\begin{ex1}\label{ex3.4}
To justify the Theorem~$\ref{th4}$ we use the same distortion function $q_{\theta}(u)$ as an Example~$\ref{ex3.3}$. Now let us consider the inverted exponential distribution defined by 
	$F(x, b) = e^{-\frac{b}{x}}$ for $x > 0$ and $b > 0$, which satisfy conditions $(i)$ and $(ii)$ of Theorem $\ref{th4}.$
	Now consider $\boldsymbol{b}=(b_0,b_1,b_2,b_3)=(0.05,0.05,0.04,0.02)$ and $\boldsymbol{b^*}=(b_0^*,b_1^*,b_2^*,b_3^*)=(0.06,0.05,0.03,0.02)$ showing that $\boldsymbol{b}\preceq_m \boldsymbol{b^*}.$ Considering $\theta_1=10$ and $\theta_2=7,$ Figure $\ref{Fig3.4}$ shows that $X_c$ is greater than $X_c^*$ in $hr$ ordering. Here substitution $x=-\ln y$, $0<y<1$, is used to capture whole real line.
\end{ex1}

%\begin{r1}
%\textcolor{blue}{It can be noticed that the condition ($ii$) of Theorem \ref{th3} (and also of Theorem \ref{th4}), that  $\frac{(1-u)q_{\theta}'(u)}{q_{\theta}(u)}$ is decreasing (increasing) in $\theta$, implies that $q_{\theta}(u)$ is log-concave (log-convex) in $\theta$. So, it can be said that, while estimating $\theta$ through MLE, log-concave (or log-convex) $q_{\theta}(u)$ with respect to $\theta$ will ensure well-behaved likelihood function, from which it is possible to get more reliable and numerically stable estimate of $\theta$. Other than $q_{\theta}(u)$ given in Example \ref{ex2a} (or of Example \ref{ex3.4}), the same function of a $3$-out-of-$4$ system with Frank copula or of a $\min\left(X_{2:4}, X_4\right)$ system having Gumbel or Frank copula satisfy condition ($ii$) of Theorem \ref{th3} (and of Theorem \ref{th4}).}
%\end{r1}

\begin{figure}[H]
			%\centering
			\begin{minipage}[]{0.48\linewidth}
				\includegraphics[height=4.0 cm, width=7.9 cm]{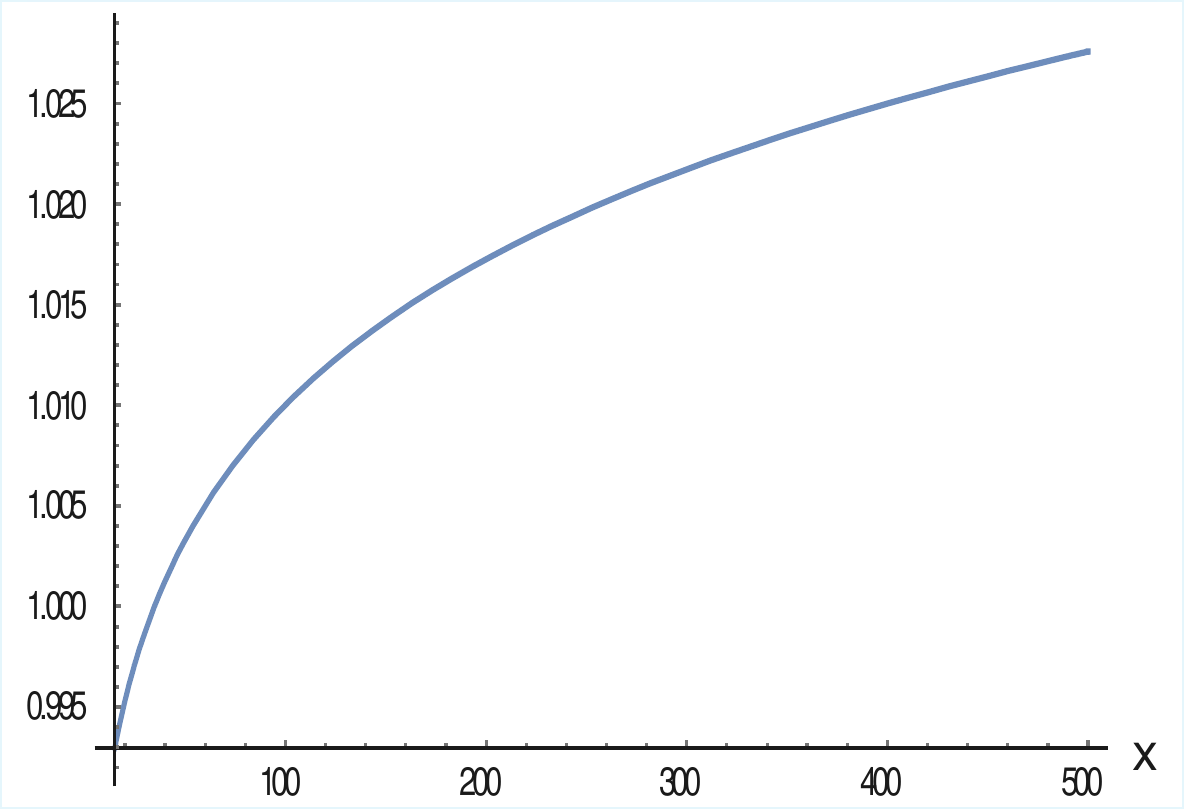}\\
				\centering{\small{$\left(a\right)$ Curve of $\frac{F_c^*(y)}{F_c(y)}$}}
			\end{minipage}
			\quad
			\begin{minipage}[]{0.48\linewidth}
				\includegraphics[height=4.0 cm, width=7.9 cm]{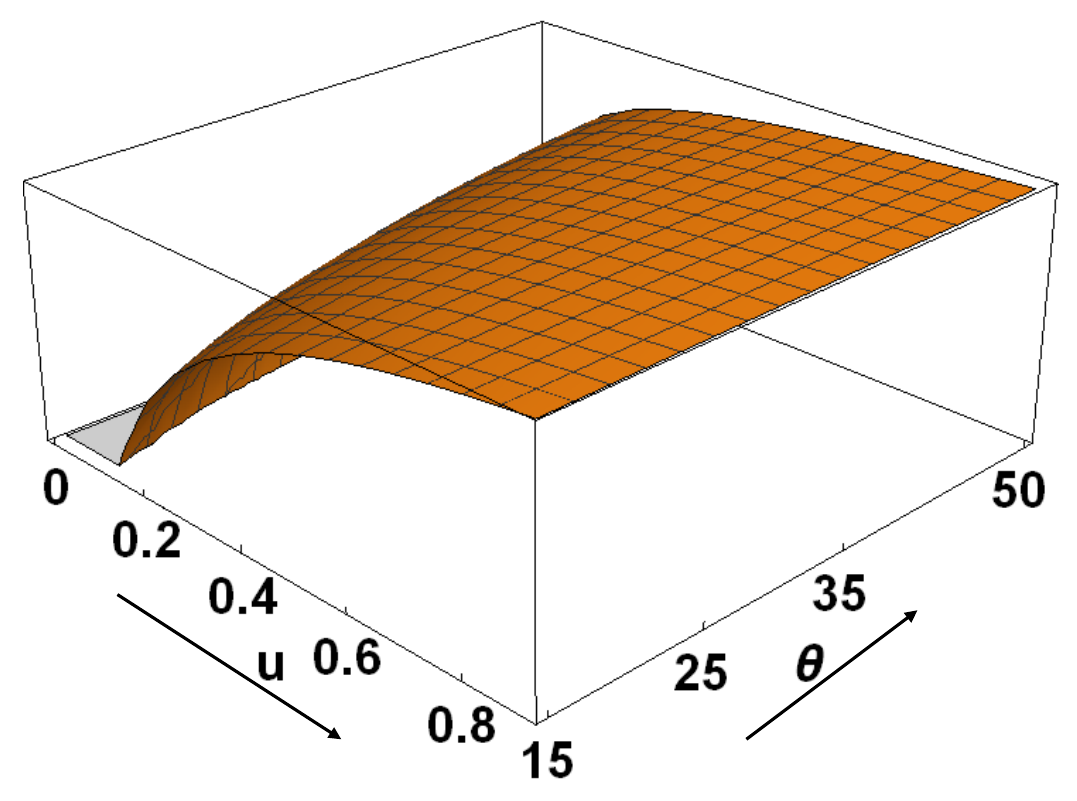}\\
				\centering{\small{ $\left(b\right)$ Curve of $\frac{(1-u)q'_{\theta}(u)}{1-q_{\theta}(u)}$}}
			\end{minipage}\caption{Curves for Example \ref{ex5}}\label{fig3}
		\end{figure}

\begin{comment}
	
\begin{ex1}
	
To justify Theorem~\ref{th4}, we also consider the inverse exponential distribution defined by 
	 $F(x, b) = e^{-\frac{\sqrt{b}}{x}}$ for $x > 0$ and $b > 0$, and use the same distortion function $q_{\theta}(u)$ as in Example~\ref{ex2}. Both the distribution function and the distortion function satisfy all the conditions of the theorem.
	
\end{ex1}
\end{comment}
The $rh$ ordering helps compare the early-life reliability of two systems. A higher $rh$ rate implies that an item is more prone to early failure. 
Theorems $\ref{th5}$ and $\ref{th6}$ exhibit sufficient conditions for $rh$ ordering between $X_c$ and $X_c^*$ and determine the best allocation choice of redundancies. The sufficient conditions can be interpreted in the same manner as those in Theorems $\ref{th3}$ and $\ref{th4},$ and are illustrated by Examples \ref{ex5} and \ref{ex6}. 
	\begin{t1}\label{th5}
		Let us consider two coherent systems  having models $\left(F, \boldsymbol{b},q_{\theta_1}\right)$ with lifetime $X_c$ and model $\left(F,\boldsymbol{b^*},q_{\theta_2}\right)$ with lifetime $X_c^*$. For $\theta_1\leq \theta_2$  if  
		\begin{itemize}
			\item [i)] $F(t;b)$ is log-concave in $b$,
			\item [ii)] $\tilde{h}(t;b)$ is convex (concave)in $b$ and 
			\item [iii)] for all $u\geq 0,$ $\frac{(1-u)q_{\theta}'(u)}{1-q_{\theta}(u)}$ is increasing (decreasing) in $u$ and increasing (decreasing) in $\theta$
		\end{itemize}
		then $\boldsymbol{b}\preceq_{m} \boldsymbol{b^*}$ implies $X_c\leq_{rh} (\geq_{rh})X_c^*$.
	\end{t1}
			\begin{ex1}	\label{ex5}
		Consider a $3$-out-of-$4$ coherent system where the components are dependent by the Gumbel copula. Then $q_{\theta}(u)=4e^{-(3(-\ln u)^{\theta})^{\frac{1}{\theta}}}-3e^{-(4(-\ln u)^{\theta})^{\frac{1}{\theta}}}$ where $\theta \in [1,\infty)$, and the ratio $\frac{q_{\theta}'(u)(1-u)}{1-q_{\theta}(u)}$ is monotonically increasing in $u$ and $\theta\in[15,\infty)$ (See Figure~$\ref{fig3}(b)$). Now, let $F(x,b)=1-\left(\frac{b}{x}\right)^{1.5},\; x\geq b$, which represents cdf of the Pareto (Type I) distribution. This distribution function satisfies conditions (i) and (ii) of Theorem \ref{th5}. So, considering  $\boldsymbol{b}=(0.5,0.5,0.4,0.2)$ and $\boldsymbol{b^*}=(0.6,0.5,0.3,0.2)$ and taking $\theta_1=20$ and $\theta_2=25$, it can be written that  $\boldsymbol{b}\preceq_m \boldsymbol{b^*}$ and also Figure $\ref{fig3}(a)$ illustrates that $X_c\preceq_{rh} X_c^*.$ 
			\end{ex1}

	The proof of $\ref{th6}$ is similar to Theorem $\ref{th5}$ and hence is omitted.
		
		\begin{t1}\label{th6}
		Let two coherent system are having  model $\left(F, \boldsymbol{b},q_{\theta_1}\right)$ with lifetime $X_c$ and model $\left(F,\boldsymbol{b^*},q_{\theta_2}\right)$ with lifetime $X_c^*$. For $\theta_1\geq\theta_2$  if  
		\begin{itemize}
			\item [i)] $F(t;b)$ is log-convex in $b$,
			\item [ii)] $\tilde{h}(t;b)$ is concave (convex) in $b$ and 
			\item [iii)] for all $u\geq 0,$ $\frac{(1-u)q_{\theta}'(u)}{1-q_{\theta}(u)}$ is increasing (decreasing) in $u$ and increasing (decreasing) in $\theta$ then
			$\boldsymbol{b}\preceq_m \boldsymbol{b^*}$ implies $X_c\geq_{rh} (\leq_{rh}) X_c^*.$
		\end{itemize}
		
		\end{t1}
		
\begin{figure}[H]
			\centering
			\includegraphics[height=4.0 cm, width=7.9 cm]{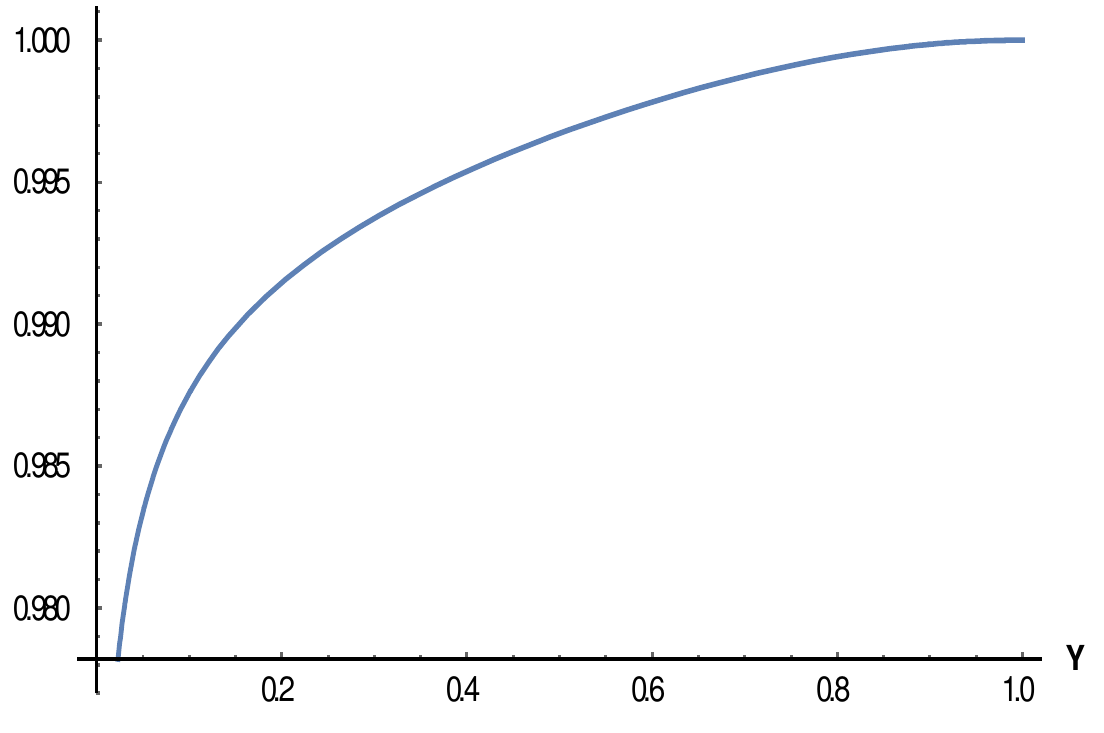}
			\caption{Curve of $\frac{F_c^*(y)}{F_c(y)}$}\label{Fig3.6}
		\end{figure}

			\begin{ex1}	\label{ex6}
				
		To illustrate Theorem~\ref{th6}, we use the same distortion function $q_{\theta}(u)$ as in Example~\ref{ex5}. Now, let $F(x; b) = \left(1-e^{-x}\right)^{\sqrt{b}}$,~for $x \geq 0,\; b \geq 0$ which represents cdf of the generalized exponential distribution, and satisfies conditions (i) and (ii) of Theorem \ref{th6}. So, considering  $\boldsymbol{b}=(0.5,0.5,0.4,0.2)$ and $\boldsymbol{b^*}=(0.6,0.5,0.3,0.2)$ and taking $\theta_1=20$ and $\theta_2=15$, it can be shown that $\boldsymbol{b}\preceq_m \boldsymbol{b^*}$ and also Figure $\ref{Fig3.6}$ illustrates that $X_c\succeq_{rh} X_c^*.$ Here substitution $x=-\ln y$, $0<y\leq 1$ is used to capture the whole real line.
		\end{ex1}
		
		%\begin{r1}
			%\textcolor{blue}{Other than $q_{\theta}(u)$ given in Example~\ref{ex5} (or Example~\ref{ex6}), 
			%the distortion function of a $2$-out-of-$3$ system with a Gumbel copula 
			%satisfies condition~(iii) of Theorem~\ref{th5} (and also of Theorem~\ref{th6}).}
			
		%\end{r1}
		
		\begin{comment}
		\begin{ex1}
			\textcolor{red}{To justify Theorem~\ref{th6}, we again consider the generalized exponential distribution} function $F(x; b) = \left(1-e^{-x}\right)^{\sqrt{b}}$ for $x \geq 0,\; b \geq 0$, and use the same distortion function $q_{\theta}(u)$ as in Example~\ref{ex5}, both of which satisfy all the conditions of the theorem.
			
		\end{ex1}
		\end{comment}
	Now the question arises, whether Theorem $\ref{th3}-\ref{th6}$ can be improved to the strongest $lr$ ordering or not. Theorem $\ref{th7}$ provides  sufficient conditions for which a coherent system with CLR can be shown to perform better than another similar system with respect to $lr$ ordering. The result is a significant improvement upon the few existing results in the present context. \\
The following definition of decreasing reversed hazard rate (DRHR) aging class is used in the next theorem and can be obtained in \cite{ah}.
		 \begin{d1} 
		 Any nonnegative absolutely continuous random variable $X$ is said to to be in DRHR aging class if the reversed hazard rate function of $X$, $\tilde{h}_X(t)$, is decreasing in t for all $t\geq 0.$
		 		 \end{d1} 
		
	\begin{t1}\label{th7}
		Let us consider two coherent systems with redundancy at component level having model $\left(F, \boldsymbol{b},q_{\theta_1}\right)$ with lifetime $X_c$ and model $\left(F,\boldsymbol{b^*},q_{\theta_2}\right)$ with lifetime $X_c^*$. For $\theta_1\geq(\leq)\theta_2$ and for all $u\geq 0$ if 
		\begin{itemize}
			
			\item [i)] $F(t;b)$ is log-concave in $b$, and in DRHR aging class,
			\item [ii)] $\tilde{h}(t;b)$ and $\tilde{h}'(t;b)$ is convex in $b$ and
			\item[iii)] for all $u\geq 0,$  $\frac{q''_{\theta}(u)(1-u)}{q'_{\theta}(u)}-1\leq0,$ is decreasing in $u$ and is increasing (decreasing) in $\theta$
		\end{itemize}
		then  $\boldsymbol{b}\preceq_{m} \boldsymbol{b^*}$ implies $X_c\leq_{lr}X_c^*.$
	\end{t1}
Theorem \ref{th7} establishes that when two systems share the same structure and distributional form with a specific class of reversed hazard rate functions, but differ in the strength of dependence among components (through $\theta$) and the quality of redundant components (through $b_i$, $i=1,2,\cdots,m$), the system with weaker (or stronger) dependence, log-concave (or log-convex) density corresponding to the distortion function, and a majorizing parameter vector achieves larger system lifetime in the sense of the $lr$ order. The following example illustrates these sufficient conditions and identifies the optimal allocation strategy for the redundant components.
	
\begin{figure}[H]
			%\centering
			\begin{minipage}[]{0.48\linewidth}
				\includegraphics[height=4.0 cm, width=7.9 cm]{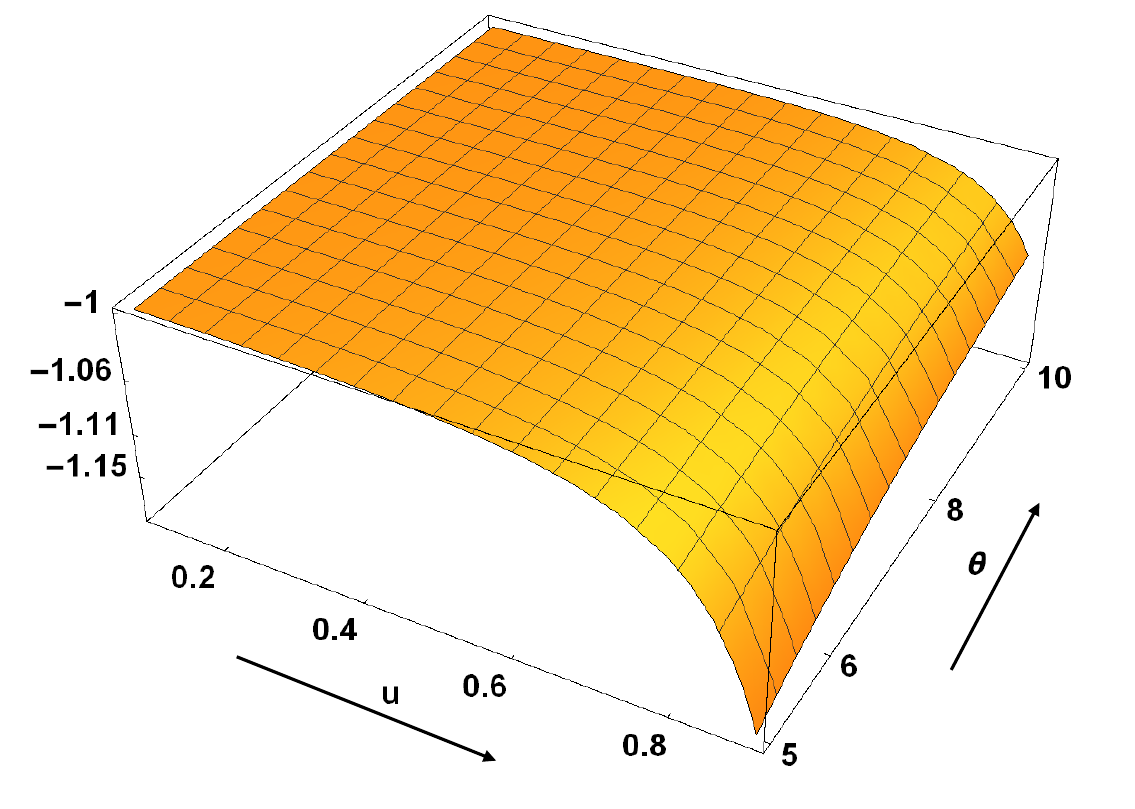}\\
				\centering{\small{$\left(a\right)$ Curve of $\Biggl\{\frac{q''_{\theta}(u)(1-u)}{q'_{\theta}(u)}-1\Biggl\}$}}
			\end{minipage}
			\quad
			\begin{minipage}[]{0.48\linewidth}
				\includegraphics[height=4.0 cm, width=7.9 cm]{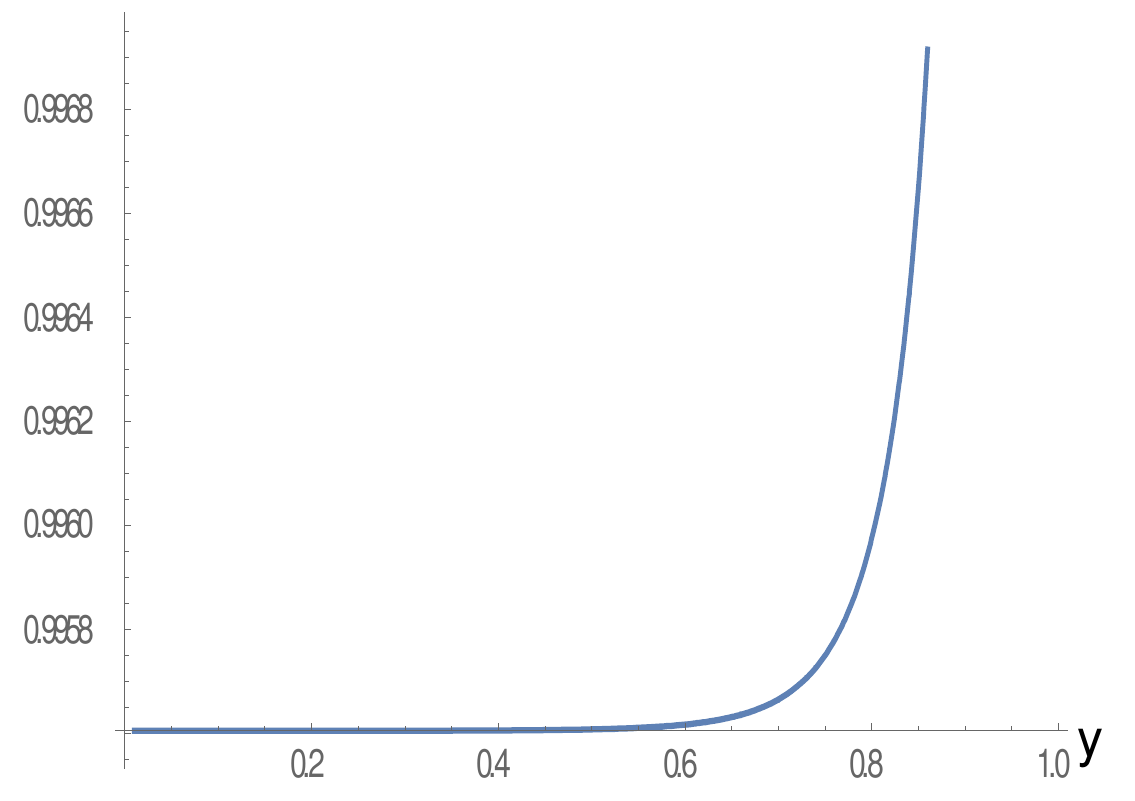}\\
				\centering{\small{ $\left(b\right)$ Curve of $\frac{f_c(y)}{f_c^*(y)}$}}
			\end{minipage}\caption{Curves for Example \ref{exam1}}\label{fig4}
		\end{figure}

	\begin{ex1}\label{exam1}
Consider a coherent system defined as $\max(X_1,\min(X_2,X_3,X_4))$, where the components are dependent on each other through the Clayton copula $C_3(u_1,u_2,u_3,u_4,\theta)=(u_1^{-\theta}+u_2^{-\theta}+u_3^{-\theta}+u_4^{-\theta}-3)^{-\frac{1}{\theta}}$. The corresponding distortion function is given by $q_{\theta}(u)=u+(3u^{-\theta}-2)^{-\frac{1}{\theta}}-(4u^{-\theta}-3)^{-\frac{1}{\theta}}$ where $\theta \in [-1, \infty)\setminus\{0\}$. Figure~\ref{fig4}(a) demonstrates that $\frac{q''_{\theta}(u)(1-u)}{q'_{\theta}(u)} \leq 1,$
and that this quantity is decreasing in $u$ while increasing in $\theta \in [5,10]$. Next, consider the inverted exponential distribution function $F(x;b_i) = e^{-\frac{b_i}{x}}$, $b_i > 0$ and $x > 0,$ which satisfies both conditions $(i)$ and $(ii)$ of the above theorem. Now, let us consider vectors $\boldsymbol{b}=(0.05,0.05,0.04,0.02)$ and $\boldsymbol{b^*}=(0.06,0.05,0.03,0.02),$ giving  $\boldsymbol{b}\preceq_m \boldsymbol{b^*}$. Now if we take $u=1-\prod_{i=0}^nF(x;b_i)$ and taking $\theta_1=7$ and $\theta_2=6$, the Figure $\ref{fig4}(b)$ ensures that $X_c\leq_{lr} X_c^*.$ Here transformation $x=-\ln y$, $0<y<1$ is used to captured whole real line.
\end{ex1}		
		
In reliability analysis, both the ordering and aging behavior of a system over its lifetime are of significant interest. Aging refers to the mathematical characterization of a system's degradation, or, in some cases, improvement over time. Various aging classes have been explored extensively in the literature. The following theorem shows that, under a specific reversed hazard rate condition on the redundant components, the lifetime of a system comprising dependent components, connected through a copula with any log-concave distortion function belongs to the $DRHR$ aging class. Example \ref{ex16} justifies the conditions. This implies that the conditional likelihood of system failure with CLR at an earlier time given that failure has already occurred before that time decreases. In other words, as time progresses, the system exhibits reduced signs of potential failure at specific earlier moments. This behavior serves as a conceptual counterpart to the Increasing Hazard Rate ($IHR$) class, offering valuable insights into failure dynamics when analyzed in reverse time.
		\begin{t1}\label{th3.8}
			Let us consider a coherent systems with redundancy at component level having model $\left(F, \boldsymbol{b},q_{\theta}\right)$ with lifetime $X_c.$ If 
			\begin{itemize}
				\item [i)] $\frac{(1-u)q'_{\theta}(u)}{1-q_{\theta}(u)}$ is increasing in $u$ and 
				\item [ii)] $\sum_{j=0}^{m}\tilde{h}(t;b_j)$ is decreasing in $t$  
			\end{itemize}
			
			then $X_c$ will be in $DRHR$ aging class.
		\end{t1}
			
\begin{figure}[H]
				%\centering
				\begin{minipage}[]{0.48\linewidth}
					\includegraphics[height=4.0 cm, width=7.9 cm]{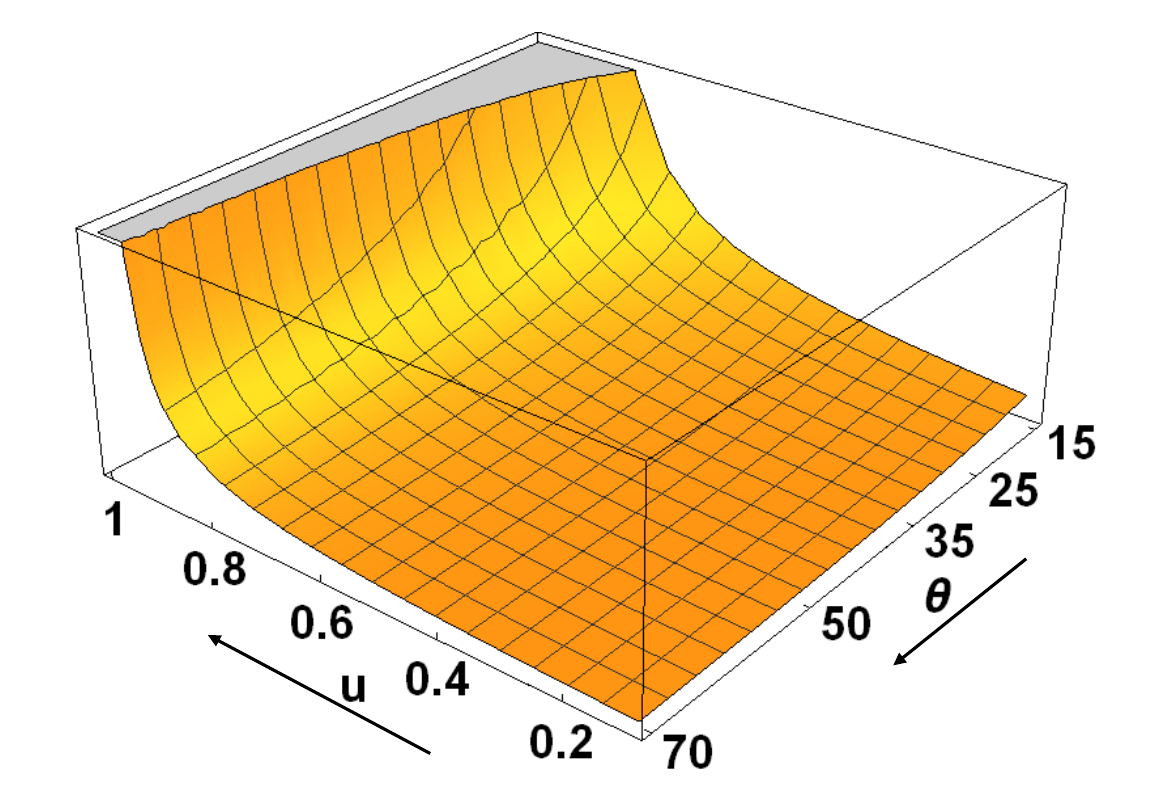}\\
					\centering{\small{$\left(a\right)$ Curve of $\frac{(1-u)q'_{\theta}(u)}{1-q_{\theta}(u)}$}}
				\end{minipage}
				\quad
				\begin{minipage}[]{0.48\linewidth}
					\includegraphics[height=4.0 cm, width=7.9 cm]{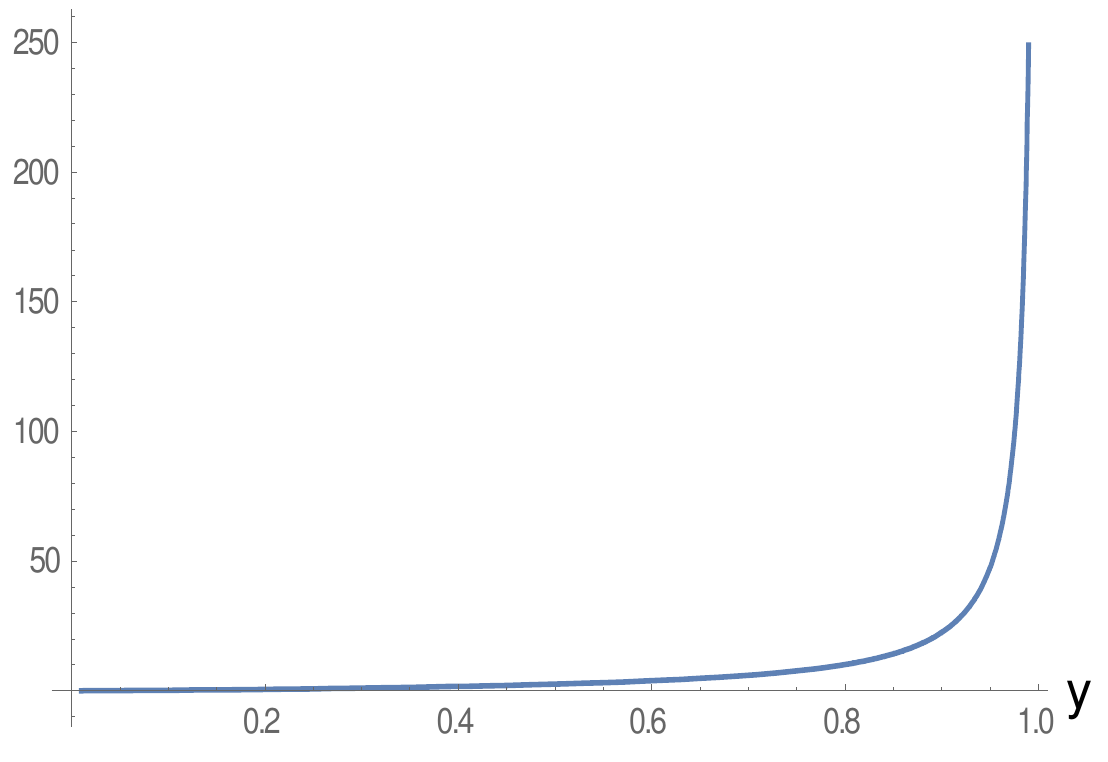}\\
					\centering{\small{ $\left(b\right)$ Curve of $\sum_{j=0}^{4} \tilde{h}(y;b_j)$}}
				\end{minipage}\caption{Curves for Example \ref{ex16}}\label{fig3.8}
			\end{figure}

	\begin{ex1}	\label{ex16}
				To justify Theorem~\ref{th3.8}, we consider the Weibull distribution function $
				F(x;b) = 1 - \exp\!\left(-x^{b}\right), 
				\; x>0,\; b>0.$
				From Figure~$\ref{fig3.8}(b)$, it is observed that $\sum_{j=0}^{4} \tilde{h}(t;b_j)$
				is decreasing in $t$  and $b=(1.6,0.5,0.3,0.2)$. Here transformation $x=-\ln y$, $0<y<1$ is used to captured whole real line. Moreover, using the distortion function 
				\[
				q_\theta(u) = 3(2u^{-\theta} - 1)^{-1/\theta} - 2(3u^{-\theta} - 2)^{-1/\theta}, 
				\quad 0 < u < 1, \, \theta > 0,
				\]
				which represents a $2$-out-of-$3$ coherent system where  components are dependent through the Clayton copula. From Figure~$\ref{fig3.8}(a)$, it can be said that for all $\theta\in[15,\infty)$, $\frac{(1-u)q'_{\theta}(u)}{1-q_{\theta}(u)}$ is increasing $u$. Hence, we observe that both conditions of Theorem~\ref{th3.8} are satisfied.
				
			\end{ex1}

	%\begin{r1}
		%\textcolor{blue}{It is to be mentioned here that, other than $q_\theta(u)$ mentioned in the above example, the same function of a $2$-out-of-$3$ system having components dependent by Gumbel copula and of a $\min\left(X_1,\max\left(X_2,X_4\right)\right)$ system with Clayton copula also satisfy the condition of Theorem \ref{th3.8}.}
	%\end{r1}	
									
	\section {Allocation of Active Redundancy at System Level}
\begin{figure}[H]
				\centering
				\includegraphics[height=4.0 cm, width=7.9 cm]{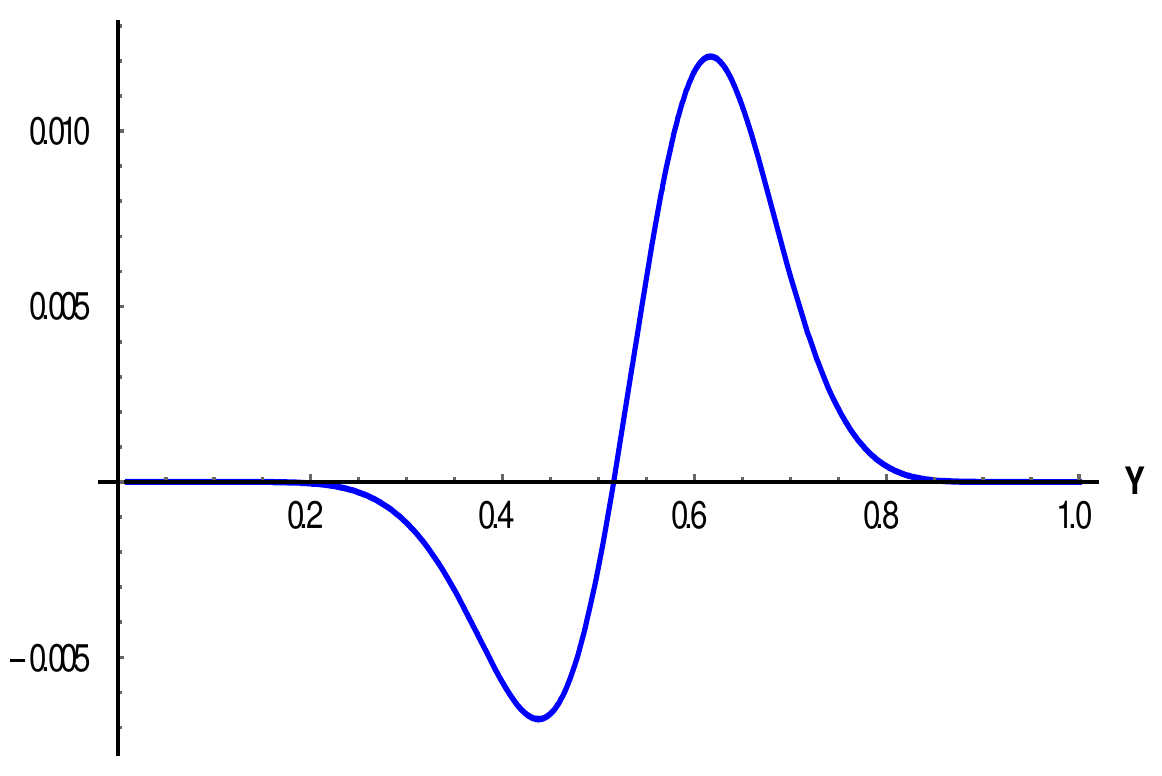}
				\caption{Curve of $\bar{F}_c(y)-\bar{F}_s(y)$}\label{BP}
	\end{figure}

	\setcounter{equation}{0}
	\hspace*{0.3in} Due to the superiority of active redundancy at component level of a system having independent components, comparisons at system level redundancy is relatively scarced in the literature. However, this superiority does not universally extend to stronger stochastic order such as $hr,$ $rhr,$ or $lr$ orders, particularly when spares are non-matching (see \cite{bo}; \cite{bri}; \cite{to}). Even, for a system having $did$ components with matching spares, $st$ ordering doesn't always hold. Example~\ref{bp} below justifies this claim. These exceptions provide strong motivation for conducting a separate and detailed investigation into the stochastic comparison of coherent systems having $did$ components with active redundancy applied at the system level.\\
			\begin{ex1}\label{bp}
				Consider a $3$-out-of-$4$ system with matching spares with dependent components through the Clayton copula. Also consider Rayleigh distribution with cdf 
				$F(t)=1-e^{-at^2}, \; t \ge 0,$ $a>0,$ and distortion function 
				\[
				q_\theta(u)=4(3u^{-\theta}-2)^{-1/\theta}-3(4u^{-\theta}-3)^{-1/\theta}, \quad 0<u<1, \, \theta>0.
				\]
				Assuming $a=1.5$ and $\theta=8.5$, it clear from the Figure~\ref{BP} that $\bar{F}_c(t)-\bar{F}_s(t)$ changes sign in its domain, showing that the BP principle does not hold hold in general even for matching spares. Here substitution $x=-\ln y$, $0<y<1$, is used, while plotting the curve.
			\end{ex1}
	
		In this section, stochastic comparison results are presented for two coherent systems of $did$ original components having cdfs $F\left(t,b_0\right)$ and $F\left(t,b^*_0\right)$ with redundancy at the system level (see Figure~\ref{fig0} (b)). Here, one original system is connected with $m$ same structured coherent systems (redundant systems) in parallel with cdf $F(t,b_j),$ $j=1,2,\ldots,m,$ while the other system is connected in a similar fashion with cdf $F\left(t,b^*_0\right).$ Let $X_s$ and $X_s^*$ denote the life times of these two systems with cdfs $F_s(t)$ and $F_s^*(t)$ respectively. Thus, using the concept of \cite{na1} as discussed in Section~\ref{sec3}, the reliability functions of these two systems can be written as
	
	\begin{equation}\label{eqs10}
		\bar{F}_s(t)=1-\prod_{j=0}^m\left(1-q_{\theta_j}(1-F(t;b_j))\right)\ {\text and }\ \bar{F}_s^*(t)=1-\prod_{j=0}^m\left(1-q_{\theta_j}(1-F(t;b_j^*))\right)
	\end{equation}
	respectively. In short, the systems are described by $(F,\boldsymbol{b},\boldsymbol{q_\theta})$ and $(F,\boldsymbol{b^*},\boldsymbol{q_\theta}),$ where $\boldsymbol{q_\theta}=(q_{\theta_0},q_{\theta_1},\cdots,q_{\theta_m})$  and the notations $\boldsymbol{b}$ and $\boldsymbol{b^*}$ are described before. Let $\mathcal{D_+} = \{(a_1,\ldots,a_n) : a_1 \geq \cdots \geq a_n\}$ and 
	$\mathcal{E_+} = \{(a_1,\ldots,a_n) : a_1 \leq \cdots \leq a_n\}$ denote the set of vectors having decreasing and increasing components, respectively. The results in this section indicate that redundancy allocation at the system level requires stricter sufficient conditions than those at the component level.

	\begin{t1}\label{th8}
			Let us consider two coherent systems with redundancy at system level having model $\left(F, \boldsymbol{b},\boldsymbol{q_\theta}\right)$ with lifetime $X_s$ and model $\left(F,\boldsymbol{b^*},\boldsymbol{q_\theta}\right)$ with lifetime $X_s^*$.  If 
			\begin{itemize}
				\item [i)] $F(t;b)$ is increasing  and log-concave in $b$, 
				\item [ii)] $\frac{(1-u)q'_{\theta}(u)}{1-q_{\theta}(u)}$ is increasing in $u$ and $\theta$, 
			\end{itemize}
		Then, for $\boldsymbol{b}, \boldsymbol{b^*} \in \mathcal{D_+}$ and $\boldsymbol{\theta} \in \mathcal{E_+}$, if $\boldsymbol{b} \preceq_m \boldsymbol{b^*}$, it follows that $X_s \leq_{st} X_s^*$. 
		\end{t1}
		
\begin{figure}[H]
				\centering
				\includegraphics[height=4.0 cm, width=7.9 cm]{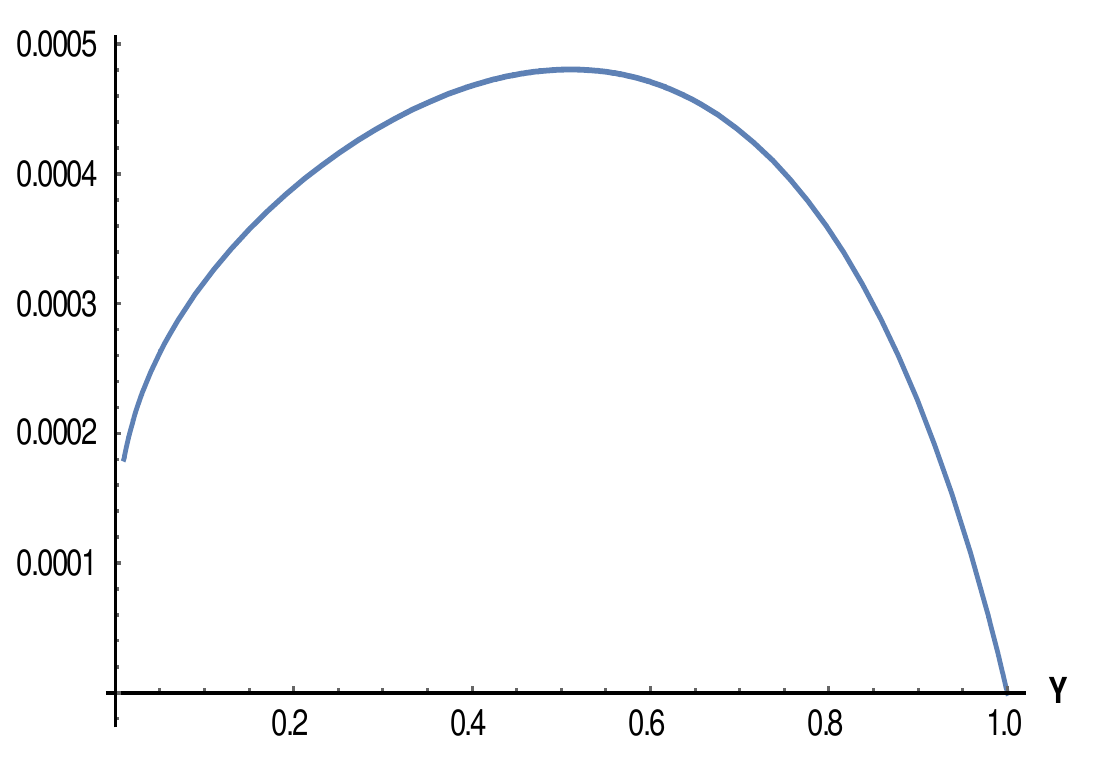}
				\caption{Curve of $\bar{F}_s^*(y)-\bar{F}_s(y)$}\label{fig5}
			\end{figure}

	Since the distortion function depends on the dependence structure of the components, the copula parameter $\boldsymbol{\theta}$ may vary across systems. The following example demonstrates the practical implications of Theorem $\ref{th8}$, showing how the effect of component quality (through the parameter $b_i$, $i=0, 1,2,\cdots,m$) influence system reliability. When the redundant systems are built from components satisfying $i)$ with a majorizing parameter vector and the dependents structure follow condition $ii)$ , the overall system lifetime becomes stochastically larger. To explain further in general, consider a fixed number of redundant systems available with varying parameters and dependency. Let us form two systems with redundancy at system level as follows: \\
					System 1: four active redundant subsystems having parameters $(b_1, b_2, b_3, b_4)$  and $(\theta_1, \theta_2, \theta_3, \theta_4).$ \\			    
					System 2: four active redundant subsystems having parameters $(b_1^{*}, b_2^{*}, b_3^{*}, b_4^{*})$ and $(\theta_1, \theta_2, \theta_3, \theta_4).$\\
							According to the sufficient conditions as explained above, the second system having a majorizing parameter vector and stronger dependence achieves a stochastically larger system lifetime. Theorem $\ref{th9}$ can be explained and proved in a similar fashion. 
		\begin{ex1}\label{ex4.2a}
		For Theorem~\ref{th8}, the same distortion function $q_{\theta}(u)$ as in Example~\ref{ex5} is used. From Figure~$\ref{fig3}(b)$ it can be checked that  $\frac{(1-u)q'_{\theta}(u)}{1-q_{\theta}(u)}$ is increasing in $u$ for $\theta\in [15,\infty).$ \\
			Let $F(x;b)=1-(1+x)^{-b}$, $b>0$ and $x\geq0$, represents the cdf of the Pareto Type-$II$ distribution and it satisfies conditions $i)$ of the aforementioned theorem.\\			
			Now, consider vectors $\boldsymbol{b}=(0.9,0.4,0.1,0.08)$, $\boldsymbol{b^*}=(0.9,0.5,0.05,0.03)$ and $\boldsymbol{\theta}=(20,21,22,23).$  It can be observed that $\boldsymbol{b}\preceq_m \boldsymbol{b^*}$  and Figure $\ref{fig5}$ illustrates that $X_s\leq_{st} X_s^*.$ Here substitution $x=-\ln y$, $0<y<1$ is used to captured the whole real line.
		\end{ex1}

				\begin{t1}\label{th9}
			Let us consider two coherent systems with redundancy at system level having model $\left(F, \boldsymbol{b},\boldsymbol{q_{\boldsymbol{\theta}}}\right)$ with lifetime $X_s$ and model $\left(F,\boldsymbol{b^*},\boldsymbol{q_{\boldsymbol{\theta}}}\right)$ with lifetime $X_s^*$.  if 
			\begin{itemize}
				\item [i)] $F(t;b)$ is decreasing  and log-convex in $b$, 
				\item [ii)] $\frac{(1-u)q'_{\theta}(u)}{1-q_{\theta}(u)}$ is decreasing in $u$ and $\theta$, 
			\end{itemize}
			Then, for $\boldsymbol{b}, \boldsymbol{b^*}\text{and}\; \boldsymbol{b^*}\in \mathcal{D_+}$  if $\boldsymbol{b} \preceq_m \boldsymbol{b^*}$, it follows that $X_s \geq_{st} X_s^*$.
		\end{t1}
		Theorem $\ref{th10}$ demonstrates that when two systems share the same structural, and distributional form with specific characteristic of hazard rate and distortion functions, but differ in the the quality of redundant systems (through $\boldsymbol{b}$), the system with a majorizing parameter vector (less dispersed allocation of component strengths) achieves larger system lifetime in terms of $rh$ ordering.
	\begin{t1}\label{th10}
		Let us consider two coherent systems with redundancy at system level having model $\left(F, \boldsymbol{b},q_{\theta}\right)$ with lifetime $X_s$ and model $\left(F,\boldsymbol{b^*},q_{\theta}\right)$ with lifetime $X_s^*$. 
		\begin{itemize}
			\item [i)] $h(t;b)$ is increasing  and convex in $b$ and 
			\item[ii)]$\frac{\bar{F}(t;b)q_{\theta}'(\bar{F}(t;b))}{1-q_{\theta}(\bar{F}(t;b))}$ is increasing and convex in $b$,
		\end{itemize}
		then  $\boldsymbol{b}\preceq_m \boldsymbol{b^*}$ implies $X_s\leq_{rh}X_s^*.$
	\end{t1}
	
\begin{figure}[ht]
			\centering
			\includegraphics[height=4.0 cm, width=7.9 cm]{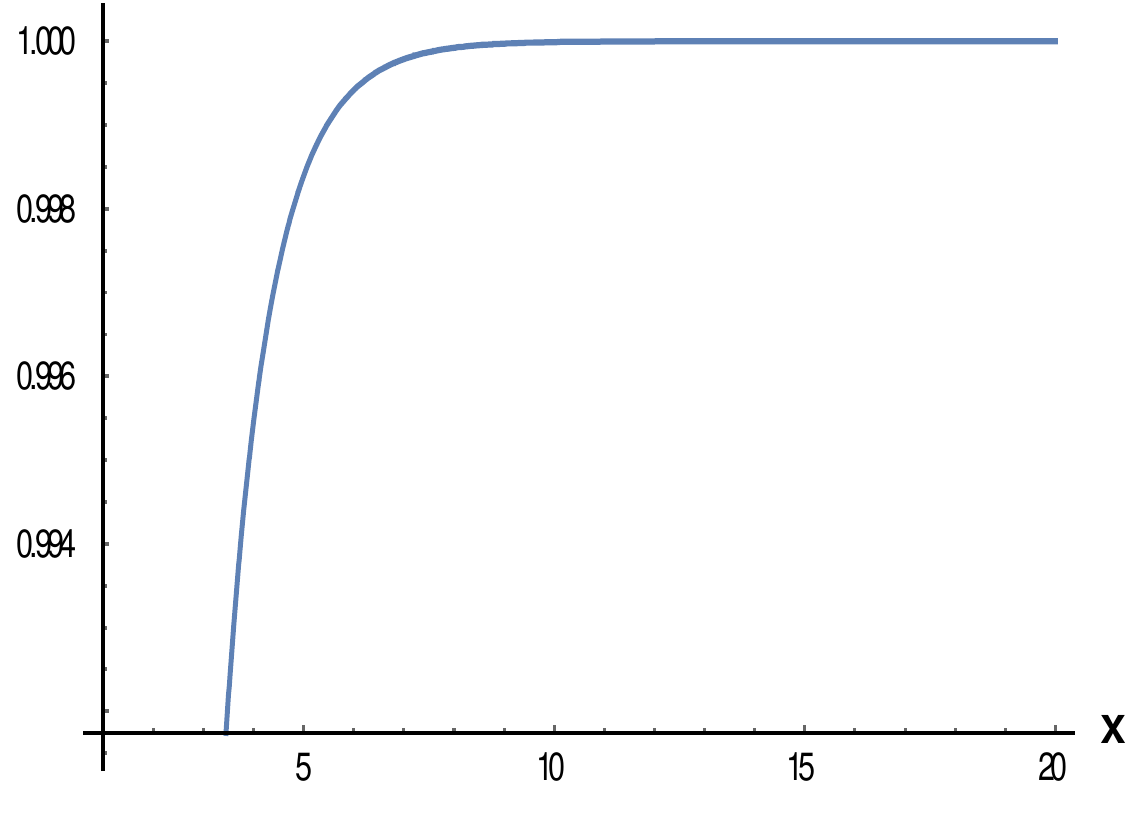}
			\caption{Curve of $\frac{F_s^*(x)} {F_s(x)}$} \label{fig6}
				%\caption{To be write}
			\end{figure}

	Next example justifies Theorem $\ref{th10}.$
	\begin{ex1}
		Consider a coherent system involving a $2$-out-of-$3$ configuration, where the components are interdependent through the Clayton copula, giving distortion function as  $$q_{\theta}(u) = 3(2u^{-\theta}-1)^{-\frac{1}{\theta}} - 2(3u^{-\theta}-2)^{-\frac{1}{\theta}}, \; \theta \in [-1, \infty)\setminus\{0\}.$$
Again, let $F(x;b)=1-e^{-(x-b)}$, $x>b$ and $b>0$, represents the cdf of the exponential distribution. So, it can be verified that $q_{\theta}(u)$ satisfies all conditions of Theorem $(\ref{th10})$  for $\theta\geq0.$\\
Now considering $\boldsymbol{b}=(0.5,0.5,0.5,0.4)$, $\boldsymbol{b^*}=(0.9,0.5,0.3,0.2)$ and $\theta=2$, giving  that $\boldsymbol{b}\preceq_m \boldsymbol{b^*}$, it can be shown from Figure $\ref{fig6}$ that $X_s\preceq_{rh} X_s^*.$
\end{ex1}			
		
The next theorem implies that if the lifetime distribution of a redundant system belongs to $DRHR$ class, the lifetime of a coherent system comprising dependent components, connected through a copula with a specific characteristic of distortion function also belongs to the DRHR aging class. Example 3.8 justifies the conditions.
\begin{comment} 
	The definition of $NBUFR$ aging class, due to \textcolor{red}{\cite{de},} is given below.
\begin{d1} 
	Any nonnegative absolutely continuous random variable $X$ having hazard rate function $h_X(t)$ is said to to be in NBUFR aging class if \textcolor{red}{ $h_X(t)\geq h_X(0)$ for all $t\geq 0.$}
\end{d1} 
\end{comment}

\begin{comment}
Theorem $\ref{the12}$ implies that for some sufficient conditions, the failure rate of a system at age $x$ with SLR is greater than or equal to the failure rate of a new system i.e., at age $0$. In other words, the failure rate of a new system with SLR is less than or equal to the failure rate of a used system.
\end{comment}

\begin{t1}\label{th4.4}
		Let us consider a coherent system with redundancy at system level having model $\left(F, \boldsymbol{b},q_{\boldsymbol{\theta}}\right)$ with lifetime $X_s$.  If 
		\begin{itemize}
			\item [i)] $\frac{uq'_{\theta}(u)}{1-q_{\theta}(u)}$ is increasing in $u$ and 
			\item [ii)] for each $b$, $F(t; b)$ belongs to the $DRHR$ aging class,
		\end{itemize}
			then $X_s$ will also be in DRHR aging class. 
		
\end{t1}
	
\begin{figure}[H]
		\centering
		\includegraphics[height=4.0 cm, width=7.9 cm]{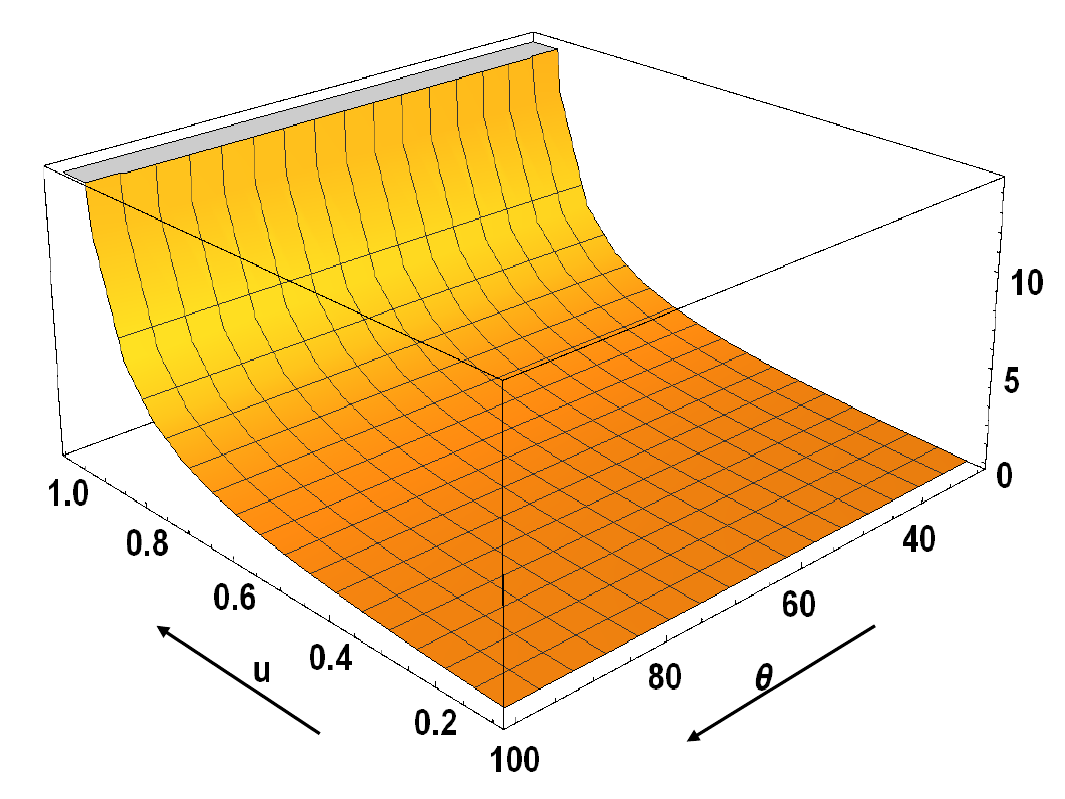}
		\caption{curve of $\frac{uq'_{\theta}(u)}{1-q_{\theta}(u)}$}\label{fig4.4}
	\end{figure}

	\begin{ex1}
		To justify Theorem~\ref{th4.4}, Consider a coherent system involving a $3$-out-of-$4$ configuration, where the components are interdependent through the Clayton copula, giving distortion function as  $$q_{\theta}(u) = 4(3u^{-\theta}-2)^{-\frac{1}{\theta}} - 3(4u^{-\theta}-3)^{-\frac{1}{\theta}}, \; \theta \in [-1, \infty)\setminus\{0\}.$$ Figure~$\ref{fig4.4}$ indicates that, $\frac{uq'_{\theta}(u)}{1-q_{\theta}(u)}$ is increasing in $u$ and $\theta.$
		Again, consider the Pareto (Type I) distribution with cdf $
		F(x;b) = 1 - \left(\frac{b}{x}\right)^{\alpha}, 
		\; x\geq b,\; \alpha>0,$ which satisfies condition $ii).$
	\end{ex1}

\noindent{\bf Acknowledgement}\\
%We thank the anonymous reviewers, the Associate Editor, and the Area Editor for their constructive comments which led to the substantive improvement of the manuscript. 
 Research scholarship grant from National Board for Higher Mathematics(NBHM)(Ref No. 0203/11/2019-R$\&$D-II/9253) is acknowledgment by Bidhan Modok. 
Shovan Chowdhury acknowledges the Financial support from Indian Institute of Management, Kozhikode, Kerala, India (Grant no. IIMK/SGRP/2024-25/08). 	
%\noindent{\bf Disclosure of Statement}\\
%The authors confirm that there are no relevant financial or non-financial competing interests to report.\\
%This research did not receive any specific grant from funding agencies in the public, commercial, or not-for-profit sectors.
	
		%\bibliographystyle{elsarticle-num}
	%\bibliography{regg}

\noindent{\bf APPENDIX}\\

\textbf{Proof of Theorem 3.1:} If $\bar{F}_c(t)$ be the s.f. of the system $\left(F, \boldsymbol{b},q_{\theta_1}\right)$, then $\bar{F}_c(t)$ can be written as 
	$$\bar{F}_c(t)=q_{\theta_1}\left(1-\prod_{j=0}^{m}F(t;b_j)\right).$$
	Let us define $\phi(\boldsymbol{b})=\prod_{j=0}^{m}F(t;b_j)$, where $\boldsymbol{b}=\left(b_0,b_1, \cdots, b_m\right).$
	Then $$\frac{\partial\phi(\boldsymbol{b})}{\partial b_i}=\left(\prod_{j=0}^{m}F(t;b_j)\right)\frac{\frac{\partial F(t;b_i)}{\partial b_i}}{F(t;b_i)}\geq 0,$$ implying that $\phi$ is increasing in each $b_i.$  Without loss of generality for $i\leq j$ let us consider $b_i\geq b_j$.
	Now, as 
	\begin{eqnarray}\label{e1}
		\frac{\partial\phi(\boldsymbol{b})}{\partial b_i}-\frac{\partial\phi(\boldsymbol{b})}{\partial b_j} &=&  \prod_{j=0}^{m}F(t;b_j)\left[\frac{\frac{\partial F(t;b_i)}{\partial b_i}}{F(t;b_i)}
		-\frac{\frac{\partial F(t;b_j)}{\partial b_j}}{F(t;b_j)}\right]\nonumber \\
		&=& \prod_{j=0}^{m}F(t;b_j)\left[\left.\frac{\partial\ln F(t;b)}{\partial b}\right|_{b=b_i}-\left.\frac{\partial\ln F(t;b)}{\partial b}\right|_{b=b_j}\right], 
	\end{eqnarray}
	Thus considering the fact  $F(t;b)$ is log-concave in $b$,  from $(\ref{e1})$ it can be concluded that 
	$$\frac{\partial\phi(\boldsymbol{b})}{\partial b_i}\leq \frac{\partial\phi(\boldsymbol{b})}{\partial b_j},$$ proving that 
	$\phi(\boldsymbol{b})$ is schur-concave in $\boldsymbol{b}$ by Theorem A.3 of \cite{ma}. Thus by Lemma $4.3$ of \cite{ku} $\boldsymbol{b}\preceq^w \boldsymbol{b}^*$ implies that $1-\prod_{j=0}^{m}F\left(t;b_j\right)\leq1-\prod_{j=0}^{m}F\left(t;b_j^*\right)$. Now, as $q_{\theta}(u)$ is increasing in $u$ and $\theta$, then for all $\theta_1\leq\theta_2$ it can be written that
	\begin{equation*}
		q_{\theta_1}\left(1-\prod_{j=0}^{m}F(t;b_j)\right)\leq q_{\theta_1}\left(1-\prod_{j=0}^{m}F(t;b_j^*)\right)\leq q_{\theta_2}\left(1-\prod_{j=0}^{m}F(t;b_j^*)\right),
	\end{equation*}
	proving the result.\hfill$\diamond$\\
	
	\textbf{Proof of Theorem 3.3:} Let us define
	$$\xi(t)=\frac{q_{\theta_2}\left(1-\prod_{j=0}^{m}F(t;b_j^*)\right)}{q_{\theta_1}\left(1-\prod_{j=0}^{m}F(t;b_j)\right)}.$$ If $\xi'(t)$ represents derivative of $\xi(t)$ with respect to $t,$ then 
	\begin{eqnarray}\label{e2}
		\xi'(t)&\overset{sign}{=}&\frac{q'_{\theta_1}\left(1-\prod_{j=0}^{m}F(t;b_j)\right) \left(\prod_{j=0}^{m}F(t;b_j)\right)\sum_{j=0}^{m}\tilde{h}(t;b_j)}{q_{\theta_1}\left(1-\prod_{j=0}^{m}F(t;b_j)\right)}\nonumber\\
		&-&\frac{q'_{\theta_2}\left(1-\prod_{j=0}^{m}F(t;b_j^*)\right)\left(\prod_{j=0}^{m}F(t;b_j^*)\right)\sum_{j=0}^{m}\tilde{h}(t;b_j^*)}{q_{\theta_2}\left(1-\prod_{j=0}^{m}F(t;b_j^*)\right)}. 
	\end{eqnarray}
	Since $F(t;b)$ is log-concave in $b$, then proceeding  same way as of Theorem $\ref{th1}$, it can be concluded that $$1-\prod_{j=0}^{m}F\left(t;b_j\right)\leq1-\prod_{j=0}^{m}F\left(t;b_j^*\right).$$
	So considering the fact that $\frac{(1-u)q_{\theta}'(u)}{q_{\theta(u)}}$ is decreasing in $u$ it can be written that
	\begin{equation}\label{e3}
		\frac{q'_{\theta}\left(1-\prod_{j=0}^{m}F(t;b_j)\right) \prod_{j=0}^{m}F(t;b_j)}{q_{\theta}\left(1-\prod_{j=0}^{m}F(t;b_j)\right)}\geq \frac{q'_{\theta}\left(1-\prod_{j=0}^{m}F(t;b_j^*)\right)\prod_{j=0}^{m}F(t;b_j^*)}{q_{\theta}\left(1-\prod_{j=0}^{m}F(t;b_j^*)\right).}
	\end{equation} 
	Again, as $\tilde{h}(t;b)$ is  concave in $b$ and $\boldsymbol{b}\preceq_{m}\boldsymbol{b^*}$, then using Lemma $\ref{l1}$ %A.2 (P. 167) of Marshall et al. (2011) it can be said that  $\boldsymbol{b}\preceq_{m}\boldsymbol{b^*}$ gives $\boldsymbol{b}\preceq^{w}\boldsymbol{b^*}$, which implies $$\left(\tilde{h}(t;b_0),\tilde{h}(t;b_1), \cdots,\tilde{h}(t;b_m)\right)\preceq^{w} \left(\tilde{h}(t;b_0^*),\tilde{h}(t;b_1^*) \cdots,\tilde{h}(t;b_m^*)\right),$$ from which it can be concluded that 
	it can be written that 
	\begin{equation}\label{e4}
		\sum_{j=0}^{m}\tilde{h}(t;b_j)\geq \sum_{j=0}^{m}\tilde{h}(t;b_j^*).
	\end{equation}
	So, from $(\ref{e3})$, $(\ref{e4})$  and  noticing the fact that $\frac{(1-u)q'_{\theta}(u)}{q_{\theta}(u)}$ is decreasing in $\theta$, it can be concluded that 
	\begin{eqnarray}\label{e5}
		\frac{q'_{\theta_1}\left(1-\prod_{j=0}^{m}F(t;b_j)\right) \prod_{j=0}^{m}F(t;b_j)\sum_{j=0}^{m}\tilde{h}(t;b_j)}{q_{\theta_1}\left(1-\prod_{j=0}^{m}F(t;b_j)\right)}\geq \frac{q'_{\theta_2}\left(1-\prod_{j=0}^{m}F(t;b_j^*)\right)\prod_{j=0}^{m}F(t;b_j^*)\sum_{j=0}^{m}\tilde{h}(t;b_j^*)}{q_{\theta_2}\left(1-\prod_{j=0}^{m}F(t;b_j^*)\right),}
	\end{eqnarray}
	proving that, $\xi(t)$ is increasing in $t$.\\
	Using the bracketed conditions it can also be proved similarly that $\xi(t)$ is decreasing in $t$.\\
	 This proves the theorem.                                 \hfill$\diamond$\\
	
	\textbf{Proof of Theorem 3.5:} Let,
	$$\xi(t)=\frac{1-q_{\theta_2}\left(1-\prod_{j=0}^{m}F(t;b_j^*)\right)}{1-q_{\theta_1}\left(1-\prod_{j=0}^{m}F(t;b_j)\right)},$$  which on differentiation with respect to $t$ gives
	\begin{eqnarray}\label{e6}
		\xi'(t)&\overset{sign}{=}&\frac{q'_{\theta_2}\left(1-\prod_{j=0}^{m}F(t;b_j^*)\right)\prod_{j=0}^{m}F(t;b_j^*)\sum_{j=0}^{m}\tilde{h}(t;b_j^*)}{1-q_{\theta_2}\left(1-\prod_{j=0}^{m}F(t;b_j^*)\right)}\nonumber\\
		&-&\frac{q'_{\theta_1}\left(1-\prod_{j=0}^{m}F(t;b_j)\right)\prod_{j=0}^{m}F(t;b_j)\sum_{j=0}^{m}\tilde{h}(t;b_j)}{1-q_{\theta_1}\left(1-\prod_{j=0}^{m}F(t;b_j)\right)}. 
	\end{eqnarray}
Now as $\boldsymbol{b}\preceq_m\boldsymbol{b^*}$ and  $\tilde{h}(t;b)$ is convex, which  in turns gives 
% gives $\boldsymbol{b}\preceq^w\boldsymbol{b^*},$ thus using Theorem A.3 of Marshall et al. (2011) and noticing the fact $\tilde{h}(t;b)$ is decreasing and convex in $b$,  it can be written that  $$\left(\tilde{h}(t;b_0),\tilde{h}(t;b_1), \cdots,\tilde{h}(t;b_m)\right)\preceq_{w} \left(\tilde{h}(t;b_0^*),\tilde{h}(t;b_1^*), \cdots,\tilde{h}(t;b_m^*)\right),$$ which  in turns gives 
	\begin{equation}\label{e7}
		\sum_{j=0}^{m}\tilde{h}(t;b_j^*)\geq \sum_{j=0}^{m}\tilde{h}(t;b_j).
	\end{equation}
	Again using condition $i)$ and proceeding in the same way as of Theorem $\ref{th1}$ it can be shown that 
	 $$1-\prod_{j=0}^{m}F\left(t;b_j\right)\leq1-\prod_{j=0}^{m}F\left(t;b_j^*\right).$$
	So, considering the fact that $\frac{(1-u)q_{\theta}'(u)}{1-q_{\theta}(u)}$ is increasing in $u$ and increasing (decreasing) in $\theta$ as well, for $\theta_1 \leq (\geq) \theta_2$ it can be concluded that  %$$1-\prod_{j=0}^{m}F(t;b_j)\leq 1-\prod_{j=0}^{m}F(t;b_j^*)$$ which implies 
	%\begin{eqnarray}\label{e8}
	%	\frac{h'_{\theta}\left(1-\prod_{j=0}^{m}F(t;b_j)\right)\left(\prod_{j=0}^{m}F(t;b_j)\right)\sum_{j=0}^{m}\tilde{h}(t;b_j)}{1-q_{\theta}\left(1-\prod_{j=0}^{m}F(t;b_j)\right)}\leq \frac{h'_{\theta}\left(1-\prod_{j=0}^{m}F(t;b_j^*)\right)\left(\prod_{j=0}^{m}F(t;b_j^*)\right)\sum_{j=0}^{m}\tilde{h}(t;b_j^*)}{1-q_{\theta}\left(1-\prod_{j=0}^{m}F(t;b_j^*)\right).} 
	%\end{eqnarray}
	%Now  from $(\ref{e8})$ and $(\ref{e9})$ and considering the fact $\frac{(1-u)q_{\theta}'(u)}{q_{\theta}(u)}$ is increasing in $\theta$,we have 
	\begin{eqnarray}\label{e8}
		0\leq\frac{q'_{\theta_1}\left(1-\prod_{j=0}^{m}F(t;b_j)\right)\prod_{j=0}^{m}F(t;b_j)}{1-q_{\theta_1}\left(1-\prod_{j=0}^{m}F(t;b_j)\right)}\leq \frac{q'_{\theta_2}\left(1-\prod_{j=0}^{m}F(t;b_j^*)\right)\prod_{j=0}^{m}F(t;b_j^*)}{1-q_{\theta_2}\left(1-\prod_{j=0}^{m}F(t;b_j^*)\right)}. 
	\end{eqnarray}
	Therefore combining $(\ref{e7})$  and $(\ref{e8})$, from $(\ref{e6})$, it is clear that $\xi(t)$ is increasing in $t.$\\
	Using the bracketed conditions it can also be proved similarly that  $\xi(t)$ is decreasing in $t.$\\ 
	This proves the result.  \hfill$\diamond$\\

\textbf{Proof of Theorem 3.7:} Let $f_c(t)$ and $f_c^*(t)$ be the probability density functions of  two systems $\left(F, \boldsymbol{b},q_{\theta_1}\right)$ and $\left(F, \boldsymbol{b^*},q_{\theta_2}\right)$ respectively, having component level redundancies. Then, 
	\begin{eqnarray*}
		f_c(t) &=& -\frac{d}{dt}\bar{F}_c(t)\\
		&=&q'_{\theta}\left(1-\prod_{j=0}^m(F(t;b_j))\right)\prod_{j=0}^m(F(t;b_j))\sum_{j=0}^m\tilde{h}(t;b_j).
	\end{eqnarray*}
	
	Let us define 
	\begin{eqnarray}\label{e9}
		\phi(t)&=& \frac{f_c(t)}{f_c^*(t)}\nonumber\\
		&=& \frac{q'_{\theta_1}\left(1-\prod_{j=0}^m(F(t;b_j))\right)\prod_{j=0}^m(F(t;b_j))\sum_{j=0}^m\tilde{h}(t;b_j)}{q'_{\theta_2}\left(1-\prod_{j=0}^m(F(t;b_j^*))\right)\prod_{j=0}^m(F(t;b_j^*))\sum_{j=0}^m\tilde{h}(t;b_j^*)}. %\ln\left(q'_{\theta}\left(1-\prod_{j=0}^m(1-F(t;b_j))\right)\right)+\ln\left(\prod_{j=0}^m(F(t;b_j))\right)+\ln\left(\sum_{j=0}^m\tilde{h}(t;b_j)\right)\\
		%&-&\ln\left(q'_{\theta}\left(1-\prod_{j=0}^m(1-F(t;b_j^*))\right)\right)-\ln\left(\prod_{j=0}^m(F(t;b_j^*))\right)-\ln\left(\sum_{j=0}^m\tilde{h}(t;b_j^*)\right).
	\end{eqnarray}	
	Considering
	$$ \psi_1(t)=\frac{q'_{\theta_1}\left(1-\prod_{j=0}^m(F(t;b_j))\right)\prod_{j=0}^m(F(t;b_j))}{q'_{\theta_2}\left(1-\prod_{j=0}^m(F(t;b_j^*))\right)\prod_{j=0}^m(F(t;b_j^*))}.
	$$
	and differentiating with respect to $t$, we have
	\begin{eqnarray}\label{e10}
		\psi_1'(t)&\overset{sign}{=}&\Biggl\{\frac{q''_{\theta_2}\left(1-\prod_{j=0}^m(F(t;b_j^*))\right)\prod_{j=0}^m(F(t;b_j^*))\sum_{j=0}^m\tilde{h}(t;b_j^*)}{q'_{\theta_2}\left(1-\prod_{j=0}^m(F(t;b_j^*))\right)}\nonumber\\&-&\frac{q''_{\theta_1}\left(1-\prod_{j=0}^m(F(t;b_j ))\right)\prod_{j=0}^m(F(t;b_j))\sum_{j=0}^m\tilde{h}(t;b_j)}{q'_{\theta_1}\left(1-\prod_{j=0}^m(F(t;b_j))\right)}\Biggl\}
		+\Bigl\{\sum_{j=0}^m\tilde{h}(t;b_j)-\sum_{j=0}^m\tilde{h}(t;b_j^*)\Bigl\}\nonumber\\
		&=& \Biggl\{ \frac{q''_{\theta_2}(v)(1-v)}{q'_{\theta_2}(v)}-1\Biggl\}\sum_{j=0}^m\tilde{h}(t;b_j^*)-\Biggl\{\frac{q''_{\theta_1}(u)(1-u)}{q'_{\theta_1}(u)}-1\Biggl\}\sum_{j=0}^m\tilde{h}(t;b_j),
	\end{eqnarray}
	where $u=1-\prod_{j=0}^m(F(t;b_j))$ and $v=1-\prod_{j=0}^m(F(t;b_j^*)).$\\
	Since $F(t;b)$ is log-concave in $b$, then proceeding  same way as of Theorem $\ref{th1}$, it can be concluded that  $1-u \geq 1-v.$
	Moreover, as  $\frac{q''_{\theta}(v)(1-v)}{q'_{\theta}(v)}$ is decreasing in $v$ and is increasing (decreasing) in $\theta$, it can be concluded that %$$\frac{q''_{\theta}(v)}{q'_{\theta}(v)}\leq \frac{q''_{\theta}(u)}{q'_{\theta}(u)}.$$
	%Now noticing the fact $\frac{d}{du}(\ln q'_{\theta}(u))$ is increasing in $\theta$, for $\theta_1\geq \theta_2$ we it can be written that 
	\begin{eqnarray}\label{e11}
		-\left\{\frac{q''_{\theta_2}(v)(1-v)}{q'_{\theta_2}(v)}-1\right\}\geq -\left\{\frac{q''_{\theta_1}(u)(1-u)}{q'_{\theta_1}(u)}-1\right\}.
	\end{eqnarray}
	Again, since $\tilde{h}(t;b)$ is convex in $b$ and $\boldsymbol{b}\preceq_m\boldsymbol{b^*}$ implies %$\boldsymbol{b}\preceq^w\boldsymbol{b^*}$ implies $$\left(\tilde{h}(t;b_0),\tilde{h}(t;b_1), \cdots,\tilde{h}(t;b_m)\right)\preceq^{w} \left(\tilde{h}(t;b_0^*),\tilde{h}(t;b_1^*), \cdots,\tilde{h}(t;b_m^*)\right),$$ which gives 
	\begin{equation}\label{e12}
		 \sum_{j=0}^{m}\tilde{h}(t;b_j^*)\geq\sum_{j=0}^{m}\tilde{h}(t;b_j).
	\end{equation}
	Thus, from $(\ref{e11})$ and $(\ref{e12})$ it can be written that 
	\begin{eqnarray}\label{e13}
		\Biggl\{\frac{q''_{\theta_2}(v)(1-v)}{q'_{\theta_2}(v)}-1\Biggl\}\sum_{j=0}^{m}\tilde{h}(t;b_j^*)\leq \Biggl\{\frac{q''_{\theta_1}(u)(1-u)}{q'_{\theta_1}(u)}-1\Biggl\}\sum_{j=0}^{m}\tilde{h}(t;b_j).
	\end{eqnarray}
	So, from $(\ref{e10})$ and $(\ref{e13})$ it can be concluded that 
	\begin{equation}\label{e14}
		\psi_1'(t)\leq 0,
	\end{equation} and hence $\psi_1(t)$ is decreasing $t\geq0$. Again let
	$$	\psi_2(t)=\frac{\sum_{j=0}^m\tilde{h}(t;b_j)}{\sum_{j=0}^m\tilde{h}(t;b_j^*)}.$$
	Therefore, for all $t\geq0$,
		\begin{eqnarray}\label{e15}
			\psi_2'(t)	&\overset{sign}{=}& \sum_{j=0}^m\tilde{h}(t;b_j^*)\sum_{j=0}^m\tilde{h}'(t;b_j)-\sum_{j=0}^m\tilde{h}(t;b_j)\sum_{j=0}^m\tilde{h}'(t;b_j^*),\nonumber\\
	          	&\overset{sign}{=}& \frac{\sum_{j=0}^m\tilde{h}'(t;b_j)}{\sum_{j=0}^m\tilde{h}(t;b_j)}-\frac{\sum_{j=0}^m\tilde{h}'(t;b_j^*)}{\sum_{j=0}^m\tilde{h}(t;b_j^*)}.
	\end{eqnarray}
	Now if we define  $\psi(\boldsymbol{b})=\frac{\sum_{j=0}^m\tilde{h}'(t;b_j)}{\sum_{j=0}^m\tilde{h}(t;b_j)}$, where $\boldsymbol{b}=(b_0,b_1,\cdots,b_m),$
	then $$\frac{\partial\psi(\boldsymbol{b})}{\partial b_k}\overset{sign}{=}\frac{\partial\tilde{h}_k'(t;b_k)}{\partial b_k}\sum_{j=0}^m\tilde{h}(t;b_j)-\frac{\partial\tilde{h}_k(t;b_j)}{\partial b_k}\sum_{j=0}^m\tilde{h}'(t;b_j).$$
	So, for $k\leq l$ assuming $b_k\geq b_l$, it can be written that $$\frac{\partial \psi(\boldsymbol{b})}{\partial b_k}-\frac{\partial \psi(\boldsymbol{b})}{\partial b_l}=\sum_{j=0}^{m}\tilde{h}_j(t;b_j)\left(\frac{\partial\tilde{h}_k'(t;b_k)}{\partial b_k}-\frac{\partial\tilde{h}_l'(t;b_l)}{\partial b_l}\right)-\sum_{j=0}^{m}\tilde{h}'_j(t;b_j)\left(\frac{\partial\tilde{h}_k(t;b_k)}{\partial b_k}-\frac{\partial\tilde{h}_l(t;b_l)}{\partial b_l}\right).$$
	Now, as $\tilde{h}_j(t;b_j)$ and $\tilde{h}'_j(t;b_j)$ are convex  and $\tilde{h}'(t;b_j)\leq 0$, then it clear from the above equation that $$\frac{\partial \psi(\boldsymbol{b})}{\partial b_k}\geq\frac{\partial \psi(\boldsymbol{b})}{\partial b_l}.$$  So, by Theorem A.3 of Marshall et al. (2011) it can be concluded that $\psi(\boldsymbol{b})$ is schur-convex in $\boldsymbol{b}$. Thus $\boldsymbol{b}\preceq_m \boldsymbol{b}^* $ implies that $\psi(\boldsymbol{b})\leq \psi(\boldsymbol{b^*}) $. So from $(\ref{e15})$ it can be concluded that 
	\begin{equation}\label{e16}
		\psi_2'(t)\leq 0,
	\end{equation}
	giving that $\psi_2(t)$ is also decreasing in $t.$
	Thus from $(\ref{e9})$ it is clear that $\phi(t)$ is decreasing in $t$, proving the result. \hfill$\diamond$\\
	
	\textbf{Proof of Theorem 3.8:} If $F_c(t)$ be the distribution function of the random variable $X_c,$ then reversed hazard rate function of $X_c$ is given by 
		\begin{eqnarray*}
			\tilde{h}_c(t) &=& \frac{d}{dt}\left(\ln (F_{c}(t))\right)\\
			&=&\frac{d}{dt}\left[\ln  \left(1-q_{\theta}\left(1-\prod_{j=0}^m F(t;b_j)\right)\right)\right]\\
			&=& \frac{q'_{\theta}\left(1-\prod_{j=0}^{m}F(t;b_j)\right)\left(\prod_{j=0}^{m}F(t;b_j)\right)\sum_{j=1}^m\tilde{h}(t;b_j)}{1-q_{\theta}\left(1-\prod_{j=0}^{m}F(t;b_j)\right)}\\
			&=&\frac{q'_{\theta}(v)(1-v)\sum_{j=0}^m\tilde{h}(t;b_j)}{1-q_{\theta}(v)}\;\text{, where} \;v=1-\prod_{j=0}^mF(t;b_j).
		\end{eqnarray*}
		Since 	$\frac{q'_{\theta}(v)(1-v)}{1-q_{\theta}(v)}$ is increasing in $v$ and $v$ is decreasing in $t$, thus $\frac{q'_{\theta}(v)(1-v)}{1-q_{\theta}(v)}$ is decreasing in $t$ and $\sum_{j=1}^m\tilde{h}(t;b_j)$ is also decreasing in $t$, proving the result. \hfill$\diamond$\\
				
		\textbf{Proof of Theorem 4.1:} Let $F_s(t)$ be the d.f. of the random variable $X_s$ and  
		\begin{equation}\label{e17}
			\Psi(\boldsymbol{b},\boldsymbol{\theta})=\ln(F_s(t))=\sum_{i=0}^{m}\ln\left(1-q_{\theta_i}(1-F(t;b_i))\right).
		\end{equation}
		Differentiating $\Psi(\boldsymbol{b},\boldsymbol{\theta})$ partially with respect to $b_j,$
		\begin{eqnarray*}
			\frac{\partial \Psi(\boldsymbol{b},\boldsymbol{\theta})}{\partial b_j} &=& \frac{q'_{\theta_j}(1-F(t;b_j)) \frac{\partial F(t;b_j)}{\partial b_j}}{1-q_{\theta_j}(1-F(t;b_j))}.\\
			&=&\frac{F(t,b_j)q'_{\theta_j}(1-F(t;b_j)) \frac{\partial(\ln{F(t;b_j)})}{\partial b_j}}{1-q_{\theta_j}(1-F(t;b_j))}
		\end{eqnarray*}
		Since $\boldsymbol{b}\in\mathcal{D_+},$ and thus for $i\leq j$,  $b_i\geq b_j$ then
		\begin{eqnarray}\label{e18}
		\frac{\partial \Psi(\boldsymbol{b},\boldsymbol{\theta})}{\partial b_i} -\frac{\partial \Psi(\boldsymbol{b},\boldsymbol{\theta})}{\partial b_j} =\frac{F(t,b_i)q'_{\theta_i}(1-F(t;b_i)) \frac{\partial(\ln{F(t;b_i)})}{\partial b_i}}{1-q_{\theta_i}(1-F(t;b_i))}-\frac{F(t,b_j)q'_{\theta_j}(1-F(t;b_j)) \frac{\partial(\ln{F(t;b_j)})}{\partial b_j}}{1-q_{\theta_j}(1-F(t;b_j))}.
		\end{eqnarray}
		So, noticing the fact  that 
		$F(t;b)$ is increasing in $b$, for $b_i\geq b_j$   it can be written that 
		\begin{equation}\label{e19}
			1-F(t;b_i)\leq 1-F(t;b_j).
		\end{equation}
Now, considering that $\frac{(1-u)q'_{\theta}(u)}{1-q_{\theta}(u)}$ is increasing in both $u$ and $\theta$, and since $\theta_i \leq \theta_j$ (as $\boldsymbol{\theta} \in \mathcal{E_+}$), it can be written that
		 \begin{equation}\label{e20}
		 	\frac{F(t,b_i)q'_{\theta_i}(1-F(t;b_i)) }{1-q_{\theta_i}(1-F(t;b_i))}\leq\frac{F(t,b_j)q'_{\theta_j}(1-F(t;b_j)) }{1-q_{\theta_j}(1-F(t;b_j))}.
		\end{equation}
		
		Again, since $F(t;b)$ is increasing and log-concave in $b$ and  $b_i\geq b_j$, then 
		\begin{eqnarray}\label{e21}
			0 \leq \frac{\partial(\ln{F(t;b_i))}}{\partial b_i}\leq \frac{\partial( \ln{F(t;b_j))}}{\partial b_j}.
		\end{eqnarray} 
		
		So, combining $(\ref{e20}),\; (\ref{e21})$   it can  be concluded from $(\ref{e18})$
		$$\frac{\partial \Psi(\boldsymbol{b},\boldsymbol{\theta})}{\partial b_i} \leq\frac{\partial \Psi(\boldsymbol{b},\boldsymbol{\theta})}{\partial b_j},$$ proving that $\Psi$ is Schur concave in $\boldsymbol{b}$ by Theorem A.3 (P. 83) of Marshall et al. (2011). Thus it can be written that 
		% Therefore $\frac{\partial F_{\boldsymbol{b},\boldsymbol{\theta}}(t)}{\partial b_k}$ is increasing in $k$and  Hence 
		\begin{eqnarray}\label{e23}
			\boldsymbol{b}\preceq_{m} \boldsymbol{b^*} \;\text{implies} \; \Psi(\boldsymbol{b},\boldsymbol{\theta})\geq \Psi(\boldsymbol{b^*},\boldsymbol{\theta}),
		\end{eqnarray} 
		proving that $X_s\leq_{st}X_s^*$
		
		\textbf{Proof of Theorem 4.3:}
	If $\tilde{h}_s(\boldsymbol{b},t)$ is reversed hazard rate function of $X_s$ then
	\begin{eqnarray*}
		\tilde{h}_s(\boldsymbol{b},t) &=& \sum_{i=0}^m\frac{q_{\theta}'\left(1-F(t;b_i)\right) f(t;b_i)}{1-q_{\theta}\left(1-F(t;b_i)\right)}\\
		&=&\sum_{i=0}^m h(t;b_i)(\bar{F}(t;b_i))\frac{q_{\theta}'(\bar{F}(t;b_i)) }{1-q_{\theta}(\bar{F}(t;b_i))}.
	\end{eqnarray*}
	So form conditions (i) and (ii) it can be concluded that $h(t;b_i)(\bar{F}(t;b_i))\frac{q_{\theta}'(\bar{F}(t;b_i))}{1-q_{\theta}(\bar{F}(t;b_i))}$ is convex in $b_i$. So, by Proposition C.1. of \cite{ma} 
	$\tilde{h}_s(\boldsymbol{b},t)$ is Schur-convex in $\boldsymbol{b}$. Hence
	$\boldsymbol{b}\preceq_m \boldsymbol{b^*} \; \text{implies}\; \tilde{h}_s(\boldsymbol{b},t)\leq\tilde{h}_s(\boldsymbol{b^*},t),$ 
proving the result. \hfill $\diamond$\\

\textbf{Proof of Theorem 4.4:} If $\tilde{h}_s(\boldsymbol{b},t)$ be the reversed hazard rate function of  random variable $X_s$.Then
	\begin{eqnarray*}
		\tilde{h}_s(\boldsymbol{b},t) &=&\frac{d}{dt}\left(\ln F_s(t;b_j)\right)\\
		&=& \frac{d}{dt}\left[\ln\left( \prod_{j=0}^m\left(1-q_{\theta}\left(1-F(t;b_j)\right)\right)\right)\right]\\
		&=& \sum_{j=0}^m\frac{q'_{\theta_j}(1-F(t;b_j))(1-F(t;b_j))h(t;b_j)}{1-q_{\theta_j}(1-F(t;b_j))}\\
		&=& \sum_{j=0}^m\frac{vq'_{\theta}(v)h(t;b_j)}{1-q_{\theta}(v)}\; \text{where}\; v=1-F(t;b_j).
	\end{eqnarray*}
	Since $\frac{vq'_{\theta}(v)}{1-q_{\theta}(v)}$ is increasing in $v$ and $v$ is decreasing in $t$, thus $\frac{vq'_{\theta}(v)}{1-q_{\theta}(v)}$ is decreasing in $t$ and also $h(t;b_j)$ is decreasing in $t$, therefore  $\tilde{h}_s(\boldsymbol{b},t)$ is decreasing in $t$, proving the theorem.\hfill $\diamond$\\

		\end{document}